\documentclass[orivec,runningheads]{llncs}

\newif\iflong\longtrue
\usepackage[dvipsnames]{xcolor}
\usepackage{tikz}
\usepackage{dsfont}
\usepackage{pgfplots}
\usepgfplotslibrary{fillbetween}
\usetikzlibrary{patterns}
\usepackage{ltl}
\usepackage{todonotes}
\usepackage[inline]{enumitem}
\usepackage{wrapfig}
\usepackage{cite}
\usepackage{subcaption}
\usepackage{stmaryrd}
\usepackage{listings}
\usepackage{multirow}
\usepackage[skins]{tcolorbox} % in the preamble
\tcbset{highlight/.style={
  colback=cyan!8,          % background
  colframe=cyan!60!black,  % border
  boxrule=0pt,
  arc=0.5mm                 % corner radius
  left=1mm, right=1mm, top=2mm, bottom=2mm     % inner padding          % border width
}}

\usepackage{amsmath}
\usepackage{caption}
\usepackage{subcaption}
\usepackage{scalerel}

\allowdisplaybreaks

\usepackage[ruled,vlined,linesnumbered]{algorithm2e}
\SetKwInOut{Input}{Input}
\SetKwInOut{Output}{Output}
\SetKwProg{Fn}{Function}{}{}
\SetNlSkip{0.3ex} 
\SetAlgoSkip{}
\usepackage[stable]{footmisc}

\usepackage{hyperref}
\hypersetup{%
	colorlinks=true,           % use colored links
	allcolors=blue!70!black,   % use dark blue for all links
	pdfstartview=Fit,          % open PDF viewer with side-pane hidden.
	breaklinks=true}
\usepackage{cleveref}
\usepackage{orcidlink}

\usetikzlibrary{arrows,backgrounds,decorations,decorations.pathmorphing,
	positioning,fit,automata,shapes,snakes,patterns,plotmarks,calc,trees}

\hypersetup{breaklinks=true}
\usepackage{soul}

\usepackage{booktabs}

\begin{document}
\newcommand{\exvars}{\mathcal{U}}
\newcommand{\envars}{\mathcal{V}}
\newcommand{\range}{\mathcal{R}}

\newcommand{\sig}{\mathcal{S}}
\newcommand{\streq}{\mathcal{F}}
\newcommand{\MOD}{\mathcal{M}}
\newcommand{\MU}{(\MOD, \vec{u})}
\newcommand{\intrv}{\mathcal{I}}

\newcommand{\cause}{\mathsf{c}}
\newcommand{\effect}{\mathsf{e}}

\newcommand{\trans}{\mathcal{T}}

\newcommand{\Trans}{\Delta}
\newcommand{\Labels}{\Lambda}
\newcommand{\Traces}{\mathsf{Tr}}

\newcommand{\mdp}{\mathbb{M}}
\newcommand{\dtmc}{\mathcal{M}}
\newcommand{\states}{S}
\newcommand{\act}{act}
\newcommand{\prob}{\mathbf{P}}
\newcommand{\init}{\iota_{init}}
\newcommand{\ap}{AP}
\newcommand{\labelfunc}{L}
\newcommand{\AW}{\mathsf{AW}}
\newcommand{\CW}{\mathsf{CW}}
\newcommand{\SE}{\mathsf{SE}}
\newcommand{\AWABS}{\widehat{\mathsf{AW}}}
\newcommand{\CWABS}{\widehat{\mathsf{CW}}}
\newcommand{\EW}{\mathsf{EqW}}
\newcommand{\ST}{\mathsf{ST}}
\newcommand{\NEW}{\mathsf{NEqW}}
\newcommand{\stuntileq}{\mathsf{stUeq}}
\newcommand{\SEABS}{\widehat{\mathsf{SE}}}

\newcommand{\pos}{\mathit{pos}}
\newcommand{\vel}{\mathit{vel}}
\newcommand{\acc}{a}
\newcommand{\halt}{\mathsf{halt}}

\newcommand{\HyperPCTL}{\textsf{\small HyperPCTL}\xspace}

\newcommand{\pac}{\mathsf{pac}}
\newcommand{\pacabs}{\mathsf{pac~abs}}
\newcommand{\lang}{\mathcal{L}}
\newcommand{\Inf}{\mathsf{Inf}}
\newcommand{\pba}{\mathcal{P}}
\newcommand{\alphabet}{\mathrm{\Sigma}}
\newcommand{\state}{s}
\newcommand{\vstate}{\vec{\state}}
\newcommand{\hstate}{\hat{\state}}
\newcommand{\hstateof}[2]{\hat{\state}_{#1}(#2)}
\newcommand{\hstates}{\hat{\states}}

\newcommand{\xpath}{\pi}
\newcommand{\Paths}[2]{\mathit{Paths}_{#1}^{#2}}
\newcommand{\fPaths}[2]{\mathit{fPaths}_{#1}^{#2}}

\newcommand{\G}{\LTLsquare}
\newcommand{\F}{\LTLdiamond}
\newcommand{\U}{\mathbin{\mathcal{U}}}
\newcommand{\X}{\LTLcircle}
\newcommand{\BigWedge}{\mathop{\scalerel*{\wedge}{\sum}}}    % match size of \sum
\newcommand{\HugeWedge}{\mathop{\scalerel*{\wedge}{\int}}}   % match size of \int (taller)
\newcommand{\MegaWedge}{\mathop{\scalerel*{\wedge}{\prod}}}

\newcommand{\naturals}{\mathbb{N}_{>0}}
\newcommand{\naturalszero}{\mathbb{N}_{\geq 0}}
\newcommand{\scheduler}{\mathfrak{s}}
\newcommand{\Schedulers}{\mathfrak{S}}
\newcommand{\sched}{\scheduler}
\newcommand{\minscheduler}{\scheduler_{*}}
\newcommand{\maxscheduler}{\scheduler^{*}}
\newcommand{\vscheduler}{\vec{\scheduler}}
\newcommand{\schedulers}[1]{\Sigma^{#1}}
\newcommand{\hscheduler}{\hat{\scheduler}}
\newcommand{\hschedulers}{\hat{\schedulers{}}}
\newcommand{\bscheduler}{\bar{\scheduler}}
\newcommand{\bschedulers}{\hat{\schedulers{}}}
\newcommand{\haltstate}{\mathcal{H}}

\newcommand{\RE}{\mathsf{RE}}
\newcommand{\REABS}{\widehat{\mathsf{RE}}}

\newcommand{\pr}{\mathbb{P}}
\renewcommand{\Pr}{\mathrm{Pr}}
\newcommand{\Cyl}{\textit{Cyl}}
\newcommand{\parallelsum}{\mathbin{\|}}
\newcommand{\emptyword}{\epsilon}

\newcommand{\plus}{+}

\newcommand{\sep}{ \ \ | \ \ }
\newcommand{\suchthat}{\,|\,}

\newcommand{\btruth}[2]{\textit{holds}_{#1,#2}}
\newcommand{\ptruth}[2]{\textit{prob}_{#1,#2}}
\newcommand{\btoi}[2]{\textit{holdsToInt}_{#1,#2}}

\renewcommand{\P}{\mathbf{P}}
\newcommand{\tpm}{\P}
\newcommand{\Act}{\mathit{Act}}
\newcommand{\supp}{\mathit{supp}}
\newcommand{\start}{\mathit{init}}
\newcommand{\tru}{\mathtt{true}}
\newcommand{\fals}{\mathtt{false}}
\newcommand{\quant}{\mathbb{Q}}

\newcommand{\modes}{Q}
\newcommand{\mode}{q}
\newcommand{\modef}{\textit{mode}}

\newcommand{\undr}[1]{\check{#1}}
\newcommand{\ovr}[1]{\hat{#1}}

\newcommand{\fail}{\mathsf{fail}}

\newcommand{\HP}{\textsf{\small HP}\xspace}

\newcommand{\action}{\alpha}

\newcommand{\EFE}{\varphi}
\newcommand{\ABS}{\emph{abstraction}}
\newcommand{\ENDO}{endogenous\xspace}
\newcommand{\EXO}{exogenous\xspace}
\newcommand{\AP}{\textsf{AP}}
\newcommand{\NMOD}{\pi \not\models \varphi}
\renewcommand{\notin}{\not\in}
\newcommand{\OVR}{O=((\mathcal{V_O},\mathcal{U_O},\mathcal{R_O}),F_O)}
\newcommand{\Iext}{\mathcal{I}_{ext}}
\newcommand{\Iint}{\mathcal{I}_{int}}

\newcommand{\FunOvrV}{\ovr{h}}
\newcommand{\FunOvrU}{\ovr{h}_{\exvars}}
\newcommand{\hintO}{\ovr{w}}

\newcommand{\Refine}{\mathsf{Refine}}
\newcommand{\FunUndV}{\undr{h}}
\newcommand{\FunUndU}{\undr{h}_{\exvars}}
\newcommand{\hintU}{\undr{w}}

\newcommand{\Rst}{\mathsf{Rst}}
\newcommand{\rst}{|}

\newcommand{\ConMod}{\mathcal{M}=((\mathcal{U}_{\mathcal{M}}, \mathcal{V}_{\mathcal{M}},\mathcal{R}_{\mathcal{M}}),F_{\mathcal{M}},I_{\mathcal{M}})}
\newcommand{\OvMod}{O=((\mathcal{U}_O, \mathcal{V}_O,\mathcal{R}_O),F_O,I_O)}
\newcommand{\UnMod}{U=((\mathcal{U}_U, \mathcal{V}_U,\mathcal{R}_U),F_U,I_U)}
\newcommand{\con}{\emph{concrete model}}

\newcommand{\dimO}{|\mathcal{R}_{\mathcal{M}}(\mathcal{U}_{\mathcal{M}})| \leq |\mathcal{R}_{\mathcal{O}}(\mathcal{U}_O)|}
\newcommand{\dimU}{|\mathcal{R}_{\mathcal{M}}(\mathcal{U}_{\mathcal{M}})| \geq |\mathcal{R}_{\mathcal{U}}(\mathcal{U}_U)|}

\newcommand{\wind}{\mathit{wind}}
\newcommand{\gravity}{\mathit{gravity}}
\newcommand{\torbu}{\mathit{turbulence}}
\newcommand{\TranExist}{\textsf{TrE}}
\newcommand{\STW}{\textsf{StW}}
\newcommand{\Pmdp}{\mathcal{P}}
\newcommand{\absor}{\textsf{Abs}}
\newcommand{\cut}{\textsf{Cuts}}

\newcommand{\speed}{\mathit{Vt}}
\newcommand{\alt}{\mathit{alt}}
\newcommand{\pit}{\theta}
\newcommand{\pitrate}{\mathit{Q}}
\newcommand{\pow}{\mathit{pow}}
\newcommand{\aoa}{\mathit{\alpha}}
\newcommand{\yaw}{\mathit{yaw}}
\newcommand{\roll}{\mathit{roll}}
\newcommand{\throt}{\delta_t}
\newcommand{\elev}{\delta_e}
\newcommand{\ail}{\delta_a}
\newcommand{\rud}{\delta_r}
\newcommand{\gforce}{\mathit{Nz}}
\newcommand{\predicate}{\mathfrak{P}}

% \definecolor{gray}{rgb}{0.5,0.5,0.5}
% \definecolor{darkgreen}{rgb}{0,0.6,0}
% \definecolor{darkblue}{rgb}{0.2, 0.2, 0.6}
% \definecolor{niceblue}{rgb}{0.16, 0.32, 0.75}
% \definecolor{niceblack}{rgb}{0.0, 0.18, 0.39}
\definecolor{prettyred}{rgb}{1.0, 0.13, 0.32}
% \definecolor{prettypink}{rgb}{0.98, 0.38, 0.5}
\definecolor{prettyblue}{rgb}{0.0, 0.30, 1.0}
% \definecolor{prettygreen}{rgb}{0.0, 0.5, 0.0}
% \definecolor{prettyorange}{rgb}{1.0, 0.33, 0.0}
% \definecolor{prettypurple}{rgb}{0.6, 0.4, 0.8}
\definecolor{prettygreen}{rgb}{0.01, 0.75, 0.24}
% \definecolor{prettyyellow}{rgb}{0.99, 0.93, 0.0}

\newcommand{\arshia}[1]{\todo[linecolor=prettyblue,backgroundcolor=prettyblue!20,bordercolor=prettyblue, inline]{Arshia: #1}}

\newcommand{\borzoo}[1]{\todo[linecolor=prettygreen,backgroundcolor=prettygreen!20,bordercolor=prettygreen,
 inline]{Borzoo: #1}}
 
 \newcommand{\ken}[1]{\todo[linecolor=prettyred,backgroundcolor=prettyred!20,bordercolor=prettyred,
 	inline]{Ken: #1}}

\newcommand{\revision}[1]{\textcolor{black}{#1}}

\newcommand{\revisionA}[1]{\textcolor{black}{#1}}

\newcommand{\vmcai}[1]{\textcolor{black}{#1}}
\newcommand{\camera}[1]{\textcolor{black}{#1}}

\newcommand{\causeclr}[1]{\textcolor{black}{#1}}
\newcommand{\effectclr}[1]{\textcolor{black}{#1}}

\title{Efficient Discovery of Actual Causality \\ in Stochastic Systems}
\author{
  Arshia Rafieioskouei\thanks{Equal contribution.}\orcidlink{0009-0002-8844-4441}\and
  Kenneth Rogale\footnotemark[1]\orcidlink{0009-0003-6213-3933}\and
  Borzoo Bonakdarpour\orcidlink{0000-0003-1800-5419}
}
%
%\authorrunning{A. Rafieioskouei et al.}
\authorrunning{A. Rafieioskouei et al.}
\institute{Michigan State University, East Lansing, MI, USA \\
  \texttt{\{rafieios, rogaleke, borzoo\}@msu.edu}}
%
           % typeset the header of the contribution
%     \author{}      
%     \institute{}
           \maketitle   
%
%     \vspace{-1cm}

\begin{abstract}
Identifying the actual cause of events in engineered systems is a fundamental challenge in system analysis. 
%
%\emph{Causality} provides a principled framework for capturing logical dependencies between events, offering insights into the underlying dynamics of a system.
%
Finding such causes becomes more challenging in the presence of noise and \vmcai{stochastic behavior} in real-world systems. %determining causal-effect relationships remains difficult. 
In this paper, we adopt the notion of \emph{probabilistic actual causality} by Fenton-Glynn, which is a probabilistic extension of Halpern and Pearl’s actual causality, and propose a novel method to formally reason about causal effect of events in \vmcai{stochastic systems}.
\vmcai{We (1) formulate the discovery of probabilistic actual causes in computing systems as an SMT problem, and (2) address the scalability challenges by introducing an \emph{abstraction-refinement} technique that improves efficiency by up to 95\%.}
We demonstrate the effectiveness of our approach through three case studies, identifying probabilistic actual causes of safety violations in (1) the Mountain Car problem, (2) the Lunar Lander benchmark, and (3) MPC controller for an F-16 autopilot simulator.

\end{abstract}

\section{Introduction}
\label{sec:intro}

Modern computing systems often operate in open-ended \vmcai{stochastic} environments.
For instance, online {\em cyber-physical systems} (CPS) operations are subject to various disturbances and noise and incur catastrophic risks with their potential misbehavior.
Such \vmcai{stochastic behavior} directly contributes to challenges in ensuring safety and developing intelligent behavior.
Thus, to prevent failures in complex or anomalous scenarios, it is imperative to reason about {\em causes} of future misbehavior rather than their {\em correlated} symptoms. 
Engineers generally build causal systems in which outputs depend only on past and present inputs.
Hence, causal inference is the natural way to explain the root causes of potential failures.

Numerous research efforts have focused on the critical challenge of identifying and explaining faults within complex systems. 
Prior work has effectively employed causal frameworks across various domains, including embedded systems \cite{gs20,ga14,gss17,wg15,gmflx13,b24,gmr10}, and automated bug localization through software trace analysis \cite{finkbeiner_et_al:LIPIcs.FSTTCS.2024.22,cdffhhms22,cffhms22,fk17,bffs23}, among several other contexts \cite{Rafieioskouei2025,7243738}. 
However, these works do not support models with \vmcai{stochastic behavior}.
When considering the probabilistic setting, analysis requires expanded notions of causality that can accommodate this \vmcai{stochastic behavior}.
\vmcai{Relatively} few works explore probabilistic causality; many that do, such as \cite{Ziemek2022,10.1007/978-3-030-99253-8_3, baier2024formalqualitymeasurespredictors}, use different causal frameworks \camera{that lack counterfactual reasoning.
By evaluating how the outcome would change in a counterfactual world where the candidate cause does not occur, while keeping the other variables in the model fixed at the values they have in the actual world, we isolate that cause’s contribution to the effect and avoid mistaking correlation for causation. 
In their absence, methods are liable to mistake correlation for causation by crediting variables that merely co-vary with the outcome or masking the causal contribution when alternative routes to the effect are active.
}

% \borzoo{this sentence is not informative. you need to say more about thier frameworks or their shortcomings.}
%
%In \cite{maldonado2025robotpouringidentifyingcauses}, the authors approach the problem from the perspective of robotic task execution.
\vmcai{While the work in \cite{maldonado2025robotpouringidentifyingcauses} also applies a probabilistic causal framework to robotics, its focus is fundamentally on task repair, identifying an effective intervention after a failure. 
This objective differs from our focus on causal discovery. 
An approach geared towards finding a sufficient repair may not need to pinpoint the precise root cause. 
Consequently, their methodology is not designed for the fine-grained analysis required to identify actual causes of the effects in stochastic systems, which is the central challenge we address.}
%
%The authors propose applying causal reasoning to repair the robotic agent’s actions. 
% \borzoo{This sentence is also not informative. There is no point of saying what others have done unless you put them in the context of what you are doing and why yours is better. This paragraph should set the stage for the next pragraph where you describe your approach.}

In this paper, we are motivated by the notion of {\em actual causality} (AC) by Halpern and Pearl~\cite{h16}, which 
focuses on the causal effect of particular events, rather than type-level causality, which attempts to make 
general statements about scientific and natural phenomena (e.g., smoking causes cancer).
AC is a formalism to deal with token-level causality, which aims to find the causal effect 
of individual events.
The original definition of AC is for deterministic systems and, hence, can only reason about causation in deterministic systems. 
For probabilistic systems, Halpern~\cite{h16} suggests ``pulling out'' all stochasticity in the exogenous variables whose values are determined by factors outside the model and defining a probability distribution over them.
This interpretation of \vmcai{stochastic behavior} limits reasoning about the causal effect of events, where the controllable actions of the system have a probabilistic nature (e.g., due to noise or imprecise measurements).
Thus, we turn to {\em probabilistic actual causality} (PAC) introduced by Fenton-Glynn~\cite{fg17} based on the concept of AC. 
Similar to the original definition of AC, PAC is based on causal settings in terms of structural equations and requires the following: (1) there exists a scenario with non-zero probability in which both the cause and its subsequent effect occur in the actual world; and (2) for any counterfactual scenario whose contingencies are identical to those of the actual world and in which the cause does not occur, the probability of the effect occurring is less than in the actual world.

Our main contribution in this paper is as follows. 
We begin with the premise that the behavior of the system under scrutiny is given as input by an acyclic discrete-time Markov chain (DTMC).
Acyclic DTMCs reflect causal models in Halpern and Pearl's definition of actual causality, which does not permit cycles. 
For instance, it is not admissible for event A to cause event B while event B causes event A.
Fenton-Glynn's causal model is derived from Bayesian networks, so the relationships between individual endogenous variables (whose values are ultimately determined by the exogenous variables) can be made explicitly probabilistic. At the level of structural equations, noise and stochasticity are captured by exogenous variables. For example, suppose we want to model a setting where event $A$ produces outcome $B$ with probability $0.5$, that is, $P(B \mid A) = 0.5$. We can write the structural equation for $B$ as $B = A \land \mathcal{U}$, where $\mathcal{U}$ is an exogenous variable. By setting $P(\mathcal{U}) = 0.5 = P(B \mid A)$, we capture the noise or stochasticity in this situation.
Our high-level goal is to design algorithms that identify the probabilistic actual cause of an effect in a DTMC.
More specifically, we aim to design algorithms that take as input (1) a DTMC $\dtmc$, and (2) a state predicate $\varphi^e$ representing the effect, and synthesize as output a state predicate $\varphi^c$ that is the probabilistic actual cause of $\varphi^e$ in $\dtmc$.
We propose two techniques:

% \begin{tcolorbox}[highlight]   % start coloured box
\begin{enumerate}
      \item Our first algorithm is based on solving an SMT instance that encodes $\dtmc$, 
        \effectclr{$\varphi^e$}, and the constraints for PAC. 
        We also encode \causeclr{$\varphi^c$} as an uninterpreted function $f$ and, hence, 
        the SMT instance is satisfiable if and only if the witness interpretation of $f$ is the actual probabilistic cause of \effectclr{$\varphi^e$} (soundness and completeness).
        
  \item Since SMT solving does not always scale to handle large models, we also develop a technique based on 
        \emph{abstraction-refinement}. In this approach, the model is first abstracted using a set of predicates. 
        If a cause is found using SMT solving on the abstract model, it is indeed an actual cause. 
        Otherwise, we refine the model to a less coarse abstraction, and the process is repeated. 
        This iterative refinement continues until a valid cause is discovered.
\end{enumerate}
% \end{tcolorbox}

We demonstrate the effectiveness of our approach through three case studies in the context of CPS with \vmcai{stochastic behavior} to identify the probabilistic actual causes of safety violations in (1) the Mountain Car problem~\cite{openaigym}, (2) the Lunar Lander benchmark~\cite{openaigym}, and (3) an MPC controller for an F-16 autopilot simulator~\cite{ARCH18:Verification_Challenges_in_F_16}.
%
%Our experiments demonstrate that the abstraction-refinement technique outperforms SMT solving on the concrete model by 70\% on average.
%
%On average, it achieves over 70\% bett performance in larger models.
%\arshia{complete}

\paragraph{Organization.} The rest of the paper is organized as follows. \Cref{sec:pre} presents the preliminary concepts while \Cref{sec:pac} formalizes the notion of PAC. The formal statement of our problem is introduced in~\Cref{sec:problem}. Our SMT-based solution and abstraction-refinement algorithm are presented in~\Cref{sec:smt,sec:abs-ref}, respectively. We evaluate our algorithms in~\Cref{sec:exp}. Related work is discussed in~\Cref{sec:related} and we conclude in~\Cref{sec:concl}. All proofs appear in the appendix.

\section{Preliminaries}
\label{sec:pre}

%\subsection{Discrete-time Markov Chains}
In this section, we present the preliminary concepts.

\begin{definition}
\label{def:dtmc}
A \emph{discrete-time Markov chain (DTMC)} is a tuple $\dtmc {=} (\states, \P, 
\AP, L)$ with the following components:

\begin{itemize}[topsep=2pt]
\item $\states$ is a nonempty finite set of {\em states};

\item $\P : \states \times \states \rightarrow [0, 1]$ is a {\em transition 
probability function} with 
$$\sum_{\state' \in \states} \P(\state, \state') =1$$
for all $\state \in \states$;

\item $\AP$ is a finite set of {\em atomic propositions}, and 

\item $L : \states \rightarrow 2^{\AP}$ is a \emph{labeling function}.\hfill\qed

\end{itemize}

\end{definition}

%We define $\pre(\state)=\{\state'\in\states\suchthat \P(\state',\state)>0\}$, $\post(\state)=\{\state'\in\states\suchthat\P(\state,\state')>0\}$.
% For each $\state\in\states$ we define $\supp(\state)=\{\state'\in\states\ |\ 
% \P(\state,\state')>0\}$. We call $\dtmc$ \emph{finite} if $S$ is finite.

An (\emph{infinite}) \emph{path of 
$\dtmc$} is an infinite sequence $\xpath =
\state_0\state_1\state_2\ldots \in \states^\omega$ of states with
$\P(\state_i, \state_{i+1}) > 0$, for all $i \geq 0$; we write
$\xpath[i]$ for $\state_i$.  Let $\Paths{\state}{\dtmc}$ denote the
set of all (infinite) paths of $\dtmc$ starting in $\state$, and
$\fPaths{\state}{\dtmc}$ denote the set of all non-empty finite prefixes of
paths from $\Paths{\state}{\dtmc}$, which we call \emph{finite
  paths}. For a finite path $\xpath =\state_0\ldots s_k \in 
\fPaths{\state_0}{\dtmc}$, $k\geq 0$, we define $|\xpath|=k + 1$.
We will also use the notations $\Paths{}{\dtmc}=\cup_{s\in
  \states}\Paths{\state}{\dtmc}$ and $\fPaths{}{\dtmc}=\cup_{s\in
  \states}\fPaths{\state}{\dtmc}$. A state $t\in \states$ is
\emph{reachable} from a state $s\in \states$ in $\dtmc$ if there
exists a finite path in $\fPaths{s}{\dtmc}$ with last state
$t$; 
we use 
$\fPaths{\state,T}{\dtmc}$ to denote the set of all finite paths from $\fPaths{\state}{\dtmc}$ with last state in $T\subseteq S$. A state $s\in
\states$ is \emph{absorbing} if and only if $\P(\state,\state)=1$, and $\P(s, t) = 0$ for all states $t \neq s$.
We label all the absorbing states with $\halt$.
\vmcai{Also, let $\haltstate$ denote the set of states labeled with $\halt$.
}

\medskip

\begin{figure}[t]
     \begin{subfigure}[b]{0.4\textwidth}
     \centering
    \scalebox{.45}{\input{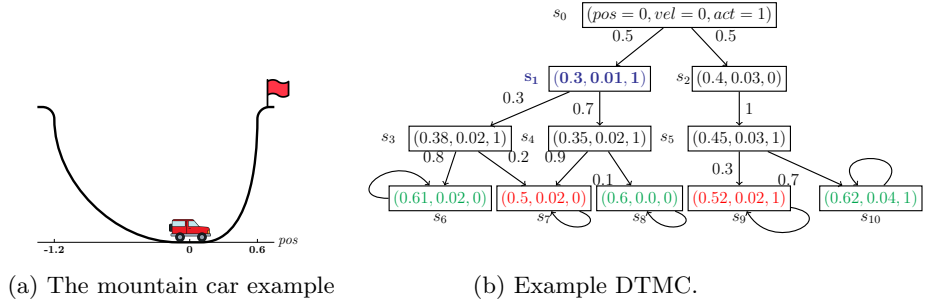}}
    \caption{The mountain car example}
    \label{fig:car_M}
\end{subfigure}
\hspace{-.5cm}
     \begin{subfigure}[b]{0.55\textwidth}        
     \centering
        \scalebox{.53}{
        \begin{tikzpicture}[->, auto, thick,node distance=3.5cm]

\large
			
			\tikzstyle{every state}=[rectangle,draw=black, text=black,minimum size=0.5cm,inner sep=0.1cm]

			\node[state]    (s0) [rectangle]             {$(pos=0,vel=0, act = 1)$};
			
			\node[state]    (s1)          [yshift=1.9cm,xshift=1.5cm,below of=s0]      {$(0.4,0.03, 0)$};
			\node[state]    (s2)            [left of=s1]    {\textcolor{Blue}{$\mathbf{(0.3,0.01, 1)}$}};

			\node[state]   (s4)          [yshift=2cm,below of=s2]      {$(0.35,0.02, 1)$};
			\node[state]   (s5)            [xshift=0cm,left of=s4]    {$(0.38,0.02, 1)$};
			
			\node[state]   (s6)          [yshift=2cm, below of=s1]      {$(0.45,0.03, 1)$};

			\node[state]    (s8)          [xshift=-0.5cm, yshift=2cm,below of=s5]      {\textcolor{Green}{$(0.61,0.02, 0)$}};
			\node[state]    (s9)            [xshift=-0.9cm,right of=s8]    {\textcolor{red}{$(0.5,0.02, 0)$}};
			\node[state]    (s10)          [yshift=2cm,xshift=1cm,below of=s4]      {\textcolor{Green}{$(0.6,0.0, 0)$}};
			
			\node[state]    (s12)          [yshift=2cm,xshift=0cm,below of=s6]      {\textcolor{red}{$(0.52,0.02, 1)$}};
			\node[state]    (s13)            [xshift=0cm,xshift=-0.2cm,right of=s12]    {\textcolor{Green}{$(0.62,0.04, 1)$}};

			\path
			
			(s0)
			edge [] node {0.5} (s1)
			edge [] node[swap] {0.5} (s2)

			(s2) edge [] node[swap] {0.7} (s4)
			edge [] node[swap] {0.3} (s5)
			
			(s1)
			edge [] node {1} (s6)

			(s5)
			edge [] node[swap] {0.8} (s8)
			edge [] node {0.2} (s9)
			(s4)
			edge [] node[swap] {0.1} (s10)
                edge [] node[swap] {0.9} (s9)
			
			(s6)
			edge [] node[swap] {0.3} (s12)
			edge [] node[swap] {0.7} (s13)
            (s13) edge [loop, looseness=5] node {} (s13)
            
            (s12) edge [out=350, in=300, looseness=3] node {} (s12)
            (s10) edge [out=340, in=300, looseness=3] node {} (s10)
            (s9) edge [out=340, in=300, looseness=3] node {} (s9)
            (s8) edge [out=180, in=130, looseness=3] node {} (s8)
            ;

                \node[left of=s0, xshift=0.5cm] {$s_0$};
                \node[left of=s1, xshift=2.1cm] {$s_2$};
                \node[left of=s2, xshift=1.8cm] {\textcolor{Blue}{$\mathbf{s_1}$}};
                \node[left of=s4, xshift=1.7cm] {$s_4$};
                \node[left of=s5, xshift=1.7cm] {$s_3$};
                \node[left of=s6, xshift=1.7cm] {$s_5$};
                
                \node[below of=s8, yshift=3cm] {$s_{6}$};
                \node[below of=s9, yshift=3cm] {$s_{7}$};
                \node[below of=s10, yshift=3cm] {$s_{8}$};
                \node[below of=s12, yshift=3cm] {$s_{9}$};
                \node[below of=s13, yshift=3cm] {$s_{10}$};

		\end{tikzpicture}}
        \caption{Example DTMC.}
        \label{fig:dtmc_example}
        \end{subfigure}
\caption{The mountain car example.}
%\vspace{-2mm}
    \end{figure}

\begin{example}
\label{ex:model}
Consider a car located in a valley and aiming to reach the top of a mountain (see 
Fig.~\ref{fig:car_M}). 
At each time step, the controller of the car determines whether to apply positive or negative acceleration to guide the car towards the mountain top. 
We model the behavior of this mountain car in the presence of uncertainty (e.g., noise) with a DTMC (see~\Cref{fig:dtmc_example}).
The tuple of three variables $(\pos, \vel, \act)$ defines the set of states in the DTMC.
The domain of $\act$ is $\{-1, 1, 0\}$, denoting negative, positive, and zero acceleration.
The stationary state $s_0$ is where $\pos=0$, $\vel=0$, and the controller decides to apply positive acceleration, i.e., $\act=1$.
Due to uncertainty, with probability 0.5, the successor states are either $s_1$, where $\pos=0.3$, $\vel=0.01$, and the controller decides to apply positive acceleration, or $s_2$, where  $\pos=0.4$, $\vel=0.03$, and the controller decides not to accelerate.
We define the safety specification of mountain car as the absorbing state that does not reach the flag, where $\pos = 0.6$, by the following predicate:
%\begin{align}
%\label{eq:fail}
$\revision{\varphi^\fail \; \triangleq \; \Big(\pos < 0.6 \wedge \halt \Big)}$.
%\end{align}
Thus, two (red) states $s_7$, $s_9$ in~\Cref{fig:dtmc_example} are labeled by $\varphi^\fail$.~\qed
\end{example}

%\subsection{Markov Decision Processes}

Markov decision processes extend DTMCs with non-deterministic choices.

\begin{definition}
A {\em Markov decision process} (\emph{MDP}) is a tuple $\mdp = (\states, \Act, 
\Pmdp, \AP, L)$ with the following components:

\begin{itemize}[topsep=2pt]

\item $\states$ is a nonempty finite set of {\em states};

\item $\Act$ is a nonempty finite set of {\em actions};

\item $\Pmdp : \states \times \Act \times \states \rightarrow [0, 1]$ is
a {\em transition probability function} such that for all $s \in 
\states$ the \emph{set of enabled actions} in $s$
\[\Act(\state)=\{\action\in\Act \suchthat \sum_{\state' \in \states} \Pmdp(\state, 
\action, \state') =1\}\]
%
% is not empty and
% %
% $$\sum_{\state' \in \states} \P(\state, \action, \state')=0$$
% %
% for all $\action \in \Act\setminus\Act(\state)$;

\item $\AP$ is a finite set of {\em atomic propositions}, and

\item $L : \states \rightarrow 2^{\AP}$ is 
a \emph{labeling function}.\hfill\qed
\end{itemize}

%We define $\pre(\state)=\{\state'\in\states\suchthat \exists 
% \alpha\in\Act(\state').\P(\state',\alpha,\state)>0\}$, 
% $\post(\state)=\{
% \state'\in\states\suchthat\exists\alpha\in\Act(\state).\P(\state,\alpha,
% \state')>0\}$.
% For each $\state\in\states$ and $\action\in\Act(\state)$ we define 
%$\supp(\state,\action)=\{\state'\in\states\ |\ \P(\state,\action,\state')>0\}$.
% We call an MDP \emph{finite} if both its state and action sets are finite.
 \end{definition}
%

%\noindent Fig.~\ref{fig:mdp} shows a simple MDP. Schedulers can be used to eliminate the non-determinism in MDPs, %inducing DTMCs with well-defined probability spaces.

%  \begin{definition}
% \label{def:scheduler}
% A {\em scheduler} for an MDP $\mdp = (\states, \Act, \P, \AP, 
% L)$ is a function $\sched: \states^+ \rightarrow \Act$, such that 
% $\sched(s_0 s_1 \cdots s_n) \in \Act(s_n)$ for all $s_0 s_1 \cdots s_n \in 
% S^+$. The path (fragment) $s_0\alpha_1 s_1 \alpha_2 s_3$ is called a 
% $\sched$-path (fragment) if $\alpha_i = \sched(s_0 \cdots s_{i-1})$ for all $i 
% > 0$. \hfill\qed
% \end{definition}
% 
% A scheduler is called {\em memoryless} if it is of the form $\sched: \states 
% \rightarrow \Act$, i.e., it picks an action only based on the current 
% state of the MDP and not the history of states. $\mdp = (\states, \Act, \P, 
% \AP, L)$ be an MDP and $\sched$ a scheduler on $\mdp$. The {\em induced Markov 
% chain} $\dtmc_\sched$ is given by $\dtmc_\sched = (\states^+, \P_\sched, \AP, 
% L')$, where for $\pi = s_0s_1 \cdots s_n$, we have $\P_\sched(\pi, \pi s_{n+1}) 
% = \P(s_n, \sched(\pi), s_{n+1})$ and $L'(\pi) = L(s_n)$.

%\borzoo{change this to simpler scheduler definition}

\begin{definition}
\label{def:scheduler}
A {\em memoryless scheduler} for an MDP $\mdp = (\states, \Act, \Pmdp, \AP, L)$ is a function $\scheduler: S \rightarrow \Act$, where $\scheduler(s) \in \Act(s)$ for all $s \in \states$.
\qed
\end{definition}
%
%Let 
%$\schedulers{\mdp}$ denote the set of all schedulers for the MDP $\mdp$. A scheduler is \emph{finite-memory} if $\modes$ is finite, \emph{memoryless} if 
%$|\modes|=1$, and \emph{non-probabilistic} if $\act(\mode,\state,\action) \in 
%\{0,1\}$ for all  $\mode\in \modes$, $s\in \states$ and $\action\in\Act$. 

\begin{definition}
\label{def:induce}
For an MDP $\mdp= (\states, \Act, \Pmdp, \AP, L)$ and a scheduler $\scheduler$ for $\mdp$, the \emph{DTMC induced by $\mdp$ and $\scheduler$} is defined as $\mdp^\scheduler = (\states, \Pmdp^\scheduler, \AP, L)$, where $\Pmdp^\scheduler(s, s') = \Pmdp(s, \scheduler(s), s')$.
\hfill\qed
\end{definition}
A state $\state'$ is \emph{reachable} from $s\in \states$ in MDP $\mdp$ if 
there exists a scheduler $\scheduler$ for $\mdp$ such that $\state'$ is 
reachable from $s$ in
$\mdp^\scheduler$. A state $s\in \states$ is \emph{absorbing} in
$\mdp$ if $s$ is absorbing in $\mdp^\scheduler$ for all schedulers
$\scheduler$ for $\mdp$.
We sometimes omit the MDP index $\mdp$ in the notations 
when it is clear from the context.
%
%\arshia{move this to sec abs-ref}
%\begin{definition}
%\label{def:schedulers}
	
Finally, for a set $B \subseteq \states$ of target states, the measure of interest is the maximum, or dually, the minimum probability of reaching a state in $B$ when starting in state $s \in \states$ among the set of all (memoryless) schedulers $\mathfrak{S}$:
\[
\mathbb{P}^{\max}(s \models \F B)  = \sup_{\scheduler \in \mathfrak{S}} ~ \mathbb{P}^{\scheduler} (s \models 
\F B)
\hspace{1cm}
\mathbb{P}^{\min}(s \models \F B)  = \inf_{\scheduler \in \mathfrak{S}} ~\mathbb{P}^{\scheduler}(s \models 
\F B).
\]

\section{Probabilistic Actual Causality}
\label{sec:pac}

As mentioned in Section~\ref{sec:intro}, we use the definition of {\em probabilistic actual 
causality} (PAC) due to Fenton-Glynn~\cite{fg17}, which is a probabilistic variation of the original 
definition of actual causality (AC) by Halpern and Pearl~\cite{h16}.
Similar to the original definition of AC~\cite{h16}, the definition of PAC in~\cite{fg17} is based on 
causal settings in terms of structural equations.
Roughly speaking, the definition of PAC requires the following: 

\begin{itemize}
	\item {\bf (PC1)} There exists a scenario with non-zero probability that the cause $\varphi^c$ and 
	the subsequent effect $\varphi^e$ both occur in the actual world. 
	
	\item {\bf (PC2)} For any counterfactual scenario whose contingencies 
	$W$ are identical to those of the actual world, 
    and in which $\varphi^c$ does not hold, the probability that $\varphi^e$ becomes true is strictly less than the probability of $\varphi^e$ occurring in the actual world.
	
\end{itemize}

In this paper, since our focus is on Markov models, we do not present the details of PAC in terms 
of structural equations.
Rather, we introduce an interpretation of PAC using the temporal logic \HyperPCTL 
with DTMC semantics~\cite{ab18}. We also do not present the full syntax and semantics of 
\HyperPCTL, as this is the only formula that we will be dealing with throughout the paper.
Let $\dtmc {=} (\states, \P, \AP, L)$ be a DTMC. The following formula expresses PAC due to 
Fenton-Glynn:
\begin{align}	\nonumber \varphi_{\pac} \triangleq \exists \sigma. \forall \sigma'. & \overbrace{\mathbb{P}\Big(\neg 
	\varphi^e_{\sigma}~\U~ (\varphi_{\sigma}^c \land \mathbb{P}_{>0}(\F 
	\varphi_{\sigma}^e))\Big)}^{\substack{\psi_{\AW}:~\text{Probability of effect } \varphi^e  \text{occurring} \\
	\text{after cause } \varphi^c \text{ in actual world } \sigma}} > 
	 \overbrace{\mathbb{P}\Big(\neg \varphi_{\sigma'} ^c \U 
	 \varphi_{\sigma'}^e\Big)}^{\substack{\psi_{\CW}:~\text{Probability of cause } \varphi^c \text{not occurring} \\ 
		\text{before effect } \varphi^e \text{ in counterfactual world } \sigma'}} \hspace{-5mm} \wedge \\
		& \label{eq:pac} \hspace{-3mm}\underbrace{\mathbb{P}_{=1} \big(\bigwedge_{a\in W} \G (a_{\sigma} \leftrightarrow 
		a_{\sigma'})\big)}_{\substack{\psi_{\SE}: \text{ Actual and counterfactual worlds} \\ \text{agree with each 
		other with respect} \\ \text{to all propositions in } W}}
\end{align}

where:

\begin{itemize}
\item $\sigma$, $\sigma'$ are two state variables that range over $\states$, 
designating the root states of the {\em actual} and {\em counterfactual} world computation trees in 
$\dtmc$, respectively.

\item $\varphi^e$ is a predicate (i.e., a Boolean combination of the propositions in $\AP$) 
expressing the effect, and $\varphi^c$ is a predicate expressing the cause. The meaning of 
$\varphi^c_\sigma$ is evaluation of formula $\varphi^c$ in the computation tree of $\dtmc$ rooted 
at state $\sigma$. Similar interpretation holds for formulas $\varphi^c_{\sigma'}$, 
$\varphi^e_{\sigma}$, and $\varphi^e_{\sigma'}$, and

\item $W \subseteq \AP$ is a subset of propositions describing all contingencies. %page 
%30 of thesis
\end{itemize}
The meaning of formula $\varphi_{\pac}$ is as follows. There exists an actual world (semantically, a 
computation tree of $\dtmc$ rooted at a state $\sigma$), such that for all counterfactual worlds (all 
computation trees rooted at a state $\sigma'$) that agree with $\sigma$ as far as propositions in 
$W$ 
are concerned (i.e., subformula $\psi_{\SE}$), and the probability of reaching the effect $\varphi^e$ after reaching the cause 
$\varphi^c$ in the actual world $\sigma$ (i.e., subformula $\psi_{\AW}$) is strictly greater than the probability of reaching the effect $\varphi^e$ without reaching the cause $\varphi^c$ in the counterfactual world $\sigma'$ (i.e., subformula $\psi_{\CW}$).

The formal semantics of $\varphi_\pac$ is based on self-composition of DTMCs.

\begin{definition}
The \emph{n-ary self-composition} of a DTMC $\dtmc = (S, \tpm, \AP, L)$ is a 
DTMC $\dtmc^n = (S^n, \tpm^n, \AP^n, L^n)$ with

\begin{itemize}

\item $S^n=S\times\ldots\times S$ is the $n$-ary Cartesian product of $S$,

\item $\tpm^n\big(s, s')=\tpm(s_1,s_1'\big)\cdot\ldots\cdot \tpm(s_n,s_n')$ for 
all $s=(s_1,\ldots,s_n)\in S^n$ and $s'=(s_1',\ldots,s_n')\in S^n$,

\item $\AP^n=\cup_{i=1}^n\AP_i$, where $\AP_i=\{a_i\,|\,a\in\AP\}$ for $i 
\in [1, n]$, and

\item $L^n(s) = \cup_{i=1}^n L_i(s_i)$ for all $s = (s_1,\ldots,s_n)\in S^n$ 
with $L_i(s_i) = \{a_i\,|\,a\in L(s_i)\}$ for $i \in [1, n]$.\hfill\qed
\end{itemize}

\end{definition}

The semantics judgment rules to evaluate formula $\varphi_\pac$ for a DTMC $\dtmc = (S, \tpm, \AP, L)$ and an $n$-tuple 
$s=(s_1,\ldots,s_n)\in S^n$ of states are the following:
\[
\begin{array}{l@{\quad}c@{\quad}l}
\dtmc,s \models \exists \sigma.\psi & \textit{iff} & 
\exists s_{n+1} \in S.\ \dtmc,(s_1,\ldots,s_n, s_{n+1}) \models \psi[\AP_{n+1}/\AP_{\sigma}]\\
\dtmc,s \models \forall \sigma.\psi & \textit{iff} & \forall s_{n+1} \in S.\ \dtmc,(s_1,\ldots,s_n, s_{n+1}) \models \psi[\AP_{n+1}/\AP_{\sigma}]\\
\dtmc,s \models a_i & \textit{iff} & a \in L(s_i)\\
\dtmc,s  \models \psi_1 \wedge \psi_2 & \textit{iff} &
\dtmc,s \models \psi_1 \textit{ and } \dtmc,s  \models \psi_2\\
\llbracket \pr(\varphi)\rrbracket_{\dtmc,s} & = & \Pr\{\pi \in \Paths{s}{\dtmc^n} \mid \dtmc,\pi \models \varphi\}\\
\dtmc,s \models p_1 > p_2 & \textit{iff} &
\llbracket p_1\rrbracket_{\dtmc,s} > \llbracket p_2 \rrbracket_{\dtmc,s}
\end{array}
\]
where $a \in \AP$ is an atomic proposition, $\sigma$ is a {\em state variable} from a countably infinite supply of variables $\mathcal{V}=\{\sigma_1,\sigma_2,\ldots\}$, $p$ is 
a \emph{probability expression}.

\camera{The satisfaction relation for \HyperPCTL path formulas is defined as follows, 
where $\pi$ is a path of $\dtmc^n$ for some $n\in\naturals$; $\psi$, $\psi_1$, 
and $\psi_2$ are \HyperPCTL state formulas:
\[
\begin{array}{l@{\quad}c@{\quad}l}
\dtmc,\pi \models \psi_1 \U \psi_2 & \textit{iff} &
\exists j \geq 0. \Big(\dtmc,\pi[j]\models\psi_2 \wedge \forall i \in [0, j). \dtmc,\pi[i] \models \psi_1\Big)\\
\end{array}
\]}
The satisfaction of formula $\varphi_\pac$ by a DTMC $\dtmc {=} (S, \tpm, \AP, L)$ is defined by:
\[
\dtmc \models \varphi_\pac \qquad \textit{iff} \qquad \dtmc,()\models\varphi_\pac
\]
\noindent where $()$ is the empty sequence of states. 

\section{Problem Statement}
\label{sec:problem}

We begin with the premise that a DTMC can be obtained by using various methods (e.g., based on learning or statistical experimental design), where each path is an experiment of the system under scrutiny, and probabilities are computed by the frequency of occurrence of states in the experiments. 
Thus, this paper is not concerned with how one obtains DTMCs.
Our goal in this paper is to design algorithms that identify the probabilistic actual causes of an effect in a DTMC.
More specifically, our algorithms take as input (1) a DTMC $\dtmc$, and (2) a predicate $\varphi^e$ and generate as output a predicate $\varphi^c$ that is the probabilistic actual cause of $\varphi^e$ in $\dtmc$.

\begin{tcolorbox}[title=Decision Problem]

Given (1) a DTMC $\dtmc$ representing a causal model, and (2) a predicate $\varphi_e$, does there exist another predicate $\varphi_c$, such that $\dtmc \models \varphi_\pac$?
\end{tcolorbox}

\begin{example}
\label{ex:smt}
Continuing with \Cref{ex:model}, we consider the DTMC shown in \Cref{fig:dtmc_example}, along with the formula $\varphi_\pac$ defined in \Cref{eq:pac} and the failure condition $\varphi^\fail$ introduced in \Cref{sec:pre}.
Suppose in~\Cref{eq:pac}, we replace $\varphi^e$ by $\varphi^{\fail}$ (i.e., the effect is failing to reach the flag). Observe that in this example $W = \{\}$. 
Now, our goal is answer the above decision problem; i.e., identifying $\varphi^c$. In this example, the answer is $\varphi^c \triangleq  \pos = 0.3 \land \vel = 0.01 \land \act = 1$:
\begin{itemize}
    \item First observe that $\mathbb{P}(\neg \varphi^e_{\sigma}~\U~ (\varphi_{\sigma}^c \land \mathbb{P}_{>0}(\F \varphi_{\sigma}^e))) = 0.5 \times 0.7 \times 0.9 + 0.5 \times 0.3 \times 0.2= 0.345$, that is the probability of reaching state $s_7$ through state $s_1$.
    %This condition requires that after satisfying the cause $\varphi^c$, the system eventually reaches the effect $\varphi^e$ with a probability greater than zero. 
    %The probability of reaching $\varphi^e$ to the t where $\pos = 0.35$ happens is 0.315, which satisfies this condition.

    \item Now, notice that $\mathbb{P}(\neg \varphi_{\sigma'} ^c \U \varphi_{\sigma'}^e) = 0.5 \times 1 \times 0.3  = 0.15$.
    
    \item Since $W = \{\}$, then $(\wedge_{a \in W} \G (a_{\sigma} \leftrightarrow a_{\sigma'}))$ trivially holds.

\end{itemize}
Consequently, since $0.345 > 0.15$, formula \Cref{eq:pac} is satisfied: predicate $\varphi^c$ is the cause for $\varphi^\fail$.~\qed

\end{example}

\section{SMT-Based Discovery of PAC}
\label{sec:smt}

An SMT decision problem consists of two main components: (1) an SMT instance, including variables, their domains, and associated symbolic structures, and (2) a set of constraints that encode the logical relationship among these variables.
\vmcai{In our SMT formulation, the objective is to symbolically identify a predicate $\varphi^c$, interpreted as the cause of the occurrence of the effect represented by the predicate $\varphi^e$. 
We introduce an uninterpreted function $f : 2^{\states} \rightarrow \{\texttt{True}, \texttt{False}\}$ which represents the cause.
For a set of states $C \subseteq \states$, we say $f(C) = \texttt{True}$ if the predicates associated with the states in $C$ only hold in the actual world but do not hold in the counterfactual world; otherwise, $f(C) = \texttt{False}$.}

\vmcai{Throughout this section, we use $p$ to denote real-valued SMT variables ranging over the interval $[0,1]$ in the SMT encodings.}
We proceed by presenting the SMT encoding of the three subformulas introduced in \Cref{eq:pac}, namely $\psi_\SE$, $\psi_\AW$, and $\psi_\CW$.

\subsubsection{Encoding equivalence of contingencies ($\boldsymbol{\psi_{\SE}}$).} 
Subformula $\psi_{\SE}$ requires that the actual and counterfactual computation trees agree on a set of all propositions $W$. 
\vmcai{However, due to potential differences in sampling frequencies, a direct state-wise comparison between paths is not meaningful. Instead, we require stutter-trace equivalence with respect to $W$.}
\vmcai{Two computation trees are stutter-trace equivalent with respect to a set of variables $W$ if all paths from the root states to the absorbing states in these trees are stutter-equivalent with respect to $W$~\cite{bk08-book}.}
\vmcai{We begin with introducing $\EW$, which verifies whether two states agree on $W$, and define it as $\EW_{\state,\state'} \triangleq \bigwedge_{a \in W} (a_{\state} \leftrightarrow a_{\state'})$.}

\vmcai{We define $\cut_m(n)$, which returns the set of all strictly increasing tuples of indices  $\langle x_0,\ldots,x_m \rangle$ taken from $\{0,\ldots,n\}$ with $x_0=0$ and $x_m=n$. 
}
\vmcai{Then, we define $\ST$, which verifies whether two paths are stutter-equivalent with respect to the variables in $W$, as:
\begin{align*}
\mathrm{\ST}_{\pi, \pi'} \triangleq
& ~~ \exists\, m\geq 1 ~.~ \exists x \in  \cut_m( |\pi| ) ~.~\exists  y \in \cut_m(|\pi'|) ~ . ~~~~~~~~~~\\
&\BigWedge_{r=0}^{m-1}\Big(
  \EW_{\pi[x_r] , \pi'[y_r]} 
  \ \land\ 
  \bigwedge_{i=x_r}^{x_{r+1}-1}\EW_{\pi[i],\,\pi[x_r]}
  \ \land
  \bigwedge_{j=y_r}^{y_{r+1}-1}\EW_{\pi'[j],\,\pi'[y_r]}
\Big)
\end{align*}
The specification above requires selecting the number of blocks $m$ such that the states in each block agree on $W$. There exists a way to partition the paths $\pi$ and $\pi'$ into $m$ blocks, where $x_r$ and $y_r$ mark the starting indices of block $r$ in $\pi$ and $\pi'$, respectively. For each block, we verify that the states at the same block index in both paths agree on $W$, and that within each block the states of each path also agree on $W$.}

\vmcai{We now define $\SE$, which captures whether two computation trees rooted in $\state$ and $\state'$ are stutter-equivalent with respect to the variables in $W$, as follows:
\[ 
\SE_{\state, \state'} \triangleq \forall \pi \in \fPaths{\state,\haltstate}{\dtmc} ~ . ~ \forall \pi' \in \fPaths{\state',\haltstate}{\dtmc} ~ . ~ \ST_{\pi,\pi'} 
\]
Here $\fPaths{\state,\haltstate}{\dtmc}$ and $\fPaths{\state',\haltstate}{\dtmc}$ are non-empty sets of paths from $\state$ and $\state'$, respectively, to states labeled by $\halt$.
}

\subsubsection{Encoding actual world computations ($\boldsymbol{\psi_{\AW}}$).} 
\vmcai{We begin by computing $p_{\RE}$, which corresponds to evaluating $\mathbb{P}(\F \varphi^e)$, that is, the probability of reaching the effect. If $\state \models \varphi^e$, then $p_{\RE_\state} = 1$; if $\state \models \neg \varphi^e \land \halt$, then $p_{\RE_\state} = 0$. Otherwise:
\[
p_{\RE_\state} = \sum_{\state' \in \states} \P(\state, \state') \cdot p_{\RE_{\state'}}
\]
Next, we compute $p_{\AW}$, which corresponds to evaluating $\psi_{\AW}$. We set $p_{\AW_\state} = 1$ if $(\state \models \varphi^c) \land (p_{\RE_\state} > 0)$. If $\big(\state \models \neg \varphi^c \land (\halt \lor \varphi^e)\big) \lor (p_{\RE_\state} = 0)$, we set $p_{\AW_\state} = 0$. Otherwise:
\[
p_{\AW_\state} = \sum_{\state' \in \states \setminus \varphi^e} \P(\state, \state') \cdot p_{\mathsf{\AW}_{\state'}}
\]
}
\subsubsection{Encoding counterfactual world computations ($\boldsymbol{\psi_{\CW}}$).} 
To reason about $\psi_\CW$, we introduce $p_{\CW_\state}$ and define it as follows: if $\state \models \varphi^e$, then $p_{\CW_\state} = 1$.
If $\state \models \neg \varphi^e \land (\halt \lor \varphi^c)$, then $p_{\CW_\state} = 0$.
Otherwise:
\[
p_{\CW_\state}  = \sum_{\state' \in \states \setminus \varphi^c} \P(\state, \state') \cdot p_{\CW_{\state'}}
\]

\subsubsection{Putting everything together.} Finally, we add the top level SMT quantification:
\[
\exists \state \in \states.~\forall \state' \in \states.~ (p_{\AW_\state} > p_{\CW_{\state'}})  \land \SE_{\state,\state'}
\]
If the SMT instance is satisfiable, then the interpretation witness of $f$ returned by the SMT solver identifies the probabilistic actual cause of $\varphi^e$.
\camera{In this formulation, we can have disjunctive causes, meaning that either $\varphi^{c_1}$ is a cause of $\varphi^e$ or $\varphi^{c_2}$ is a cause of $\varphi^e$. This is different from stating that $\varphi^{c_1} \lor \varphi^{c_2}$ is itself the cause of $\varphi^e$.}

\vmcai{It is apparent that the SMT encoding presented in this section is correct by construction, as they directly mirror the formal PAC conditions in \Cref{eq:pac}. In addition, solving the equations in this section does not require fixed-point reasoning, as our DTMCs are acyclic and have bounded depth, the recursions always terminate in absorbing states.}
\camera{The complexity of \HyperPCTL\ model checking is polynomial in the size of $\dtmc$ while being \textsf{PSPACE}-hard with respect to the number of quantifiers appearing in the formula~\cite{ab18}. The complexity of reasoning the uninterpreted function $f$ matches that of the corresponding SMT theory.
}

\section{Abstraction-Refinement for Probabilistic Causal Models }
\label{sec:abs-ref}

% \begin{wrapfigure}[13]{r}{0.4\textwidth}     
%   \vspace{-\baselineskip}                  
%   \begin{minipage}{\linewidth}            
%   \begin{algorithm}[H]
% 	\footnotesize
% 	\caption{Finding probabilistic cause using Abstraction Refinement }
% 	\label{alg:absref}
% 	\KwIn{$\dtmc$, $\varphi_e$, $\varphi_{\pac}$ , $\mathcal{P}$}
% 	\KwOut{$\varphi_c$}
%         % $\mathcal{G} \leftarrow \{\dtmc^{1}, \dots, \dtmc^n\}$\;\label{line:sub}
%         $\hat{\dtmc}_{init} \leftarrow \hat{h}(\dtmc, \mathcal{P})$\; \label{line:init_abs}
%         \While{true}{
% 		result, $\varphi_c  \leftarrow \mathsf{SMT}(\hat{\dtmc}_j, \varphi_{\pac})$\;\label{line:smt}
% 		\If {result} {
% 			\textbf{return} $\varphi_c$\;
% 		}
% 		\Else{
% 			$\hat{\dtmc}_{j+1}  \leftarrow \mathsf{Refine}(\hat{\dtmc}_j)$\;\label{line:refine}
% 		}	     
% 	}
% \end{algorithm}
%   \end{minipage}
% \end{wrapfigure}

%In this section, we introduce our iterative abstraction-refinement technique and its application in reasoning about PAC. 
%
The overall idea of our algorithm is the following steps:
%
% (see \Cref{alg:absref}):
\begin{enumerate}
% \item \textbf{Discovering subgraphs}: Identify the set of subgraphs $\{\dtmc^{1}, \dots, \dtmc^n\}$ of DTMC $\dtmc$, each representing an equivalence class of paths that are mutually stutter-trace equivalent with respect to $W$ (\Cref{line:sub}).
\item \textbf{Predicate abstraction}: Start with the concrete DTMC, $\dtmc$, and apply predicate abstraction to derive the initial abstract model, $\hat{\dtmc}$ as an MDP.
\item \textbf{SMT-Based discovery}: Identify potential causal relation within the abstract model $\hat{\dtmc}$. \vmcai{If a cause is discovered, it is confirmed as the explanation and the process terminates.}
\item \textbf{Refinement}: If no causal relationship is identified, refine $\hat{\dtmc}$ by splitting the relevant abstract state and revisit step 2.
\end{enumerate}
We explain the details of Steps 1 -- 3 in \Cref{sec:pred,sec:smt-abs,sec:ref}, respectively.

\newcommand{\eval}[1]{\llbracket #1 \rrbracket}

\subsection{Predicate Abstraction}
\label{sec:pred}
%
%Predicate abstraction is a widely used technique in abstraction-refinement algorithms \cite{10.1007/978-3-540-70545-1_16, cgjlv00}. 
%
Predicates are Boolean expressions defined over the variables, and for any such expression $\psi$, its valuation is a function $\eval{\psi}: S \rightarrow \{0, 1\}$, where $\eval{\psi}_{\state} = 1$ means state $\state$ satisfies predicate $\psi$.
%
%Predicate $\psi$ corresponds to the set of states that satisfy it, denoted by $\llbracket \psi \rrbracket$.
Following the construction in~\cite{10.1007/978-3-540-70545-1_16}, let $\predicate = \{\psi_1, \psi_2, \dots, \psi_n\}$ denote a set of predicates.
The set $\predicate$ induces a partitioning of the state space into disjoint equivalence classes based on which predicates hold.
Each equivalence class is represented as an $n$-bit vector, where the $i$-th bit indicates whether the corresponding predicate $\psi_i$ is satisfied in that class.
These bit vectors represent abstract states, which we denote by $\hat{\states}$. 
For a given abstract state $\hat{s} \in \hat{\states}$, we refer to the corresponding equivalence class as the concretization of $\hat{s}$, denoted by $\gamma(\hat{s})$. 
We define an abstraction function as $\hat{h}(s) = \big(\llbracket \psi_1 \rrbracket_{s}, \dots, \llbracket \psi_n \rrbracket_{s}\big)$. 
% \arshia{shouldnt be $\hat{h}: 2^\states \to \hat{\states}$ ?}
%
This abstraction function induces an MDP $\hat{\dtmc} = \big(\hat{\states}, \mathcal{P}, \AP, \hat{L}, \Act\big)$, where

\begin{itemize}
    \item $\Act(\hat{s}) = \{\, \action \in \gamma(\hat{s}) \mid \sum_{\hat{s}' \in \hat{\states}} \mathcal{P}(\hat{s}, \action, \hat{s}') = 1 \,\}$;

    \item $\mathcal{P}(\hat{s}, \action, \hat{s}') = \sum_{s'\in \gamma(\hat{s}')}\P(\action, s') $, and 

    \item $\hat{L}(\hat{\state}) = \bigcup_{\state \in \gamma(\hat{\state})} L(\state)$.
\end{itemize}
\vmcai{The process of state abstraction transforms a DTMC into an MDP. Nondeterminism arises when an abstract state groups together multiple concrete states with different transition probabilities. This ambiguity is resolved by interpreting the distinct transition distributions inherited from the concrete states as the set of available actions in the corresponding abstract state of the MDP.
}

\begin{lemma}
\label{lemma:1}
Let $\hat{\dtmc}$ be the abstract model obtained through predicate abstraction. If $\hat{\dtmc}$ satisfies a reachability property, then the concrete model $\dtmc$ also satisfies that property. \qed
\end{lemma}
The soundness of this lemma is guaranteed by the fact that the abstract model $\hat{\dtmc}$ simulates the concrete model $\dtmc$ \cite{10.1007/978-3-540-70545-1_16}.

\begin{example}
\label{ex:pred}

    Continuing~\Cref{ex:model} and the DTMC in~\Cref{fig:dtmc_example}, for the set of predicates $\predicate = \{\vel \geq 0.03, \pos \geq 0.6, \pos \geq 0.4, \pos \geq 0.3\}$, the states in the DTMC are partitioned into six abstract states. Using the abstraction function $\hat{h}$, we construct the MDP in~\Cref{fig:mdp_example}. For example, concrete state $s_7$ is mapped to abstract state $\hat{s}_3$ since $\hat{h}(s_7) = (0, 0, 1, 1)$. 
    \camera{The available actions of an abstract state are induced by the concrete states grouped into that abstract state based on $\predicate$. For instance, in~\Cref{fig:mdp_example}, $\Act(\hat{s}_1) = \{\state_1, \state_3, \state_4 \}$.}
\end{example}

Before delving into the SMT-based discovery of probabilistic actual causes in the abstract model, we first address $\psi_\SE$ in \Cref{eq:pac}.
In the abstraction-refinement technique, handling equivalence for $W$ is challenging and we propose two techniques:

\begin{enumerate}

    \item \textbf{Enumerating Subgraphs:} We begin by decomposing the original DTMC $\dtmc$ into subgraphs $\{\dtmc^{1}, \dots, \dtmc^n\}$, where each $\dtmc^i$ is a subgraph in which all paths are mutually stutter-equivalent with respect to $W$. Within each subgraph $\dtmc^i$, the objective is to identify a set of states representing the predicate $\varphi^c$. More details are provided in \Cref{app:subgraph}.
    \item \textbf{$W$-Preserving Abstraction:} In this approach, we restrict the abstraction function $\hat{h}$ to preserve the equality of computation trees with respect to variables in $W$. That is, for two concrete transitions $\P_{>0}(\state_\sigma, \state_\sigma')$ and $\P_{>0}(\state_{\sigma'}, \state_{\sigma'}')$, if (1) $\bigwedge_{a\in W} (a_{\state_\sigma} \leftrightarrow a_{\state_{\sigma'}})$ and (2) $\bigwedge_{a\in W} (a_{\state_\sigma'} \not\leftrightarrow a_{\state_{\sigma'}'})$, then we require that (1) $\bigwedge_{a\in W} (a_{\state_\sigma} \leftrightarrow a_{\hat{h}(\state_{\sigma'})})$ and (2) $\bigwedge_{a\in W} (a_{\state'_\sigma} \not\leftrightarrow a_{\hat{h}(\state'_{\sigma'})})$, \vmcai{where $\sigma$ and $\sigma'$ are state variables}. Otherwise, we will not be able to prove the soundness of the abstraction-refinement algorithm with respect to the contingencies defined by $W$. 
\end{enumerate}

\subsection{Discovery of Probabilistic Causes in Abstract Model}
\label{sec:smt-abs}
To identify probabilistic actual causes in the abstract MDP, we first need to adapt the SMT-based encoding originally designed for DTMCs, as discussed in Section~\ref{sec:smt}. 
%
%In the DTMC setting, the reachability probabilities of the effect states are deterministically computed.
%
%Due to abstraction, the model exhibits nondeterministic choices over the underlying concrete states represented by each abstract state. 
%
%As a result, 
Since we are dealing with an MDP, we compute the range of reachability probabilities, by finding the schedulers that render minimum and maximum of $\psi_\AW$ and $\varphi_\CW$, respectively:
%
%Therefore, we verify the following formula:
%
% \begin{align}
% \nonumber \hat{\varphi}_{\pac}  \triangleq \exists \hat{\sigma}. \forall \hat{\sigma}'. & \overbrace{\mathbb{P}^{\text{min}} \Big(\neg \varphi^e_{\hat{\sigma}}~\U~ (\varphi_{\hat{\sigma}}^c \land \mathbb{P}^{\text{min}}_{>0}(\F 
% 	\varphi_{\hat{\sigma}}^e))\Big)}^{\substack{\psi_\AWABS\text{:Minimum probability of effect } \varphi^e  \text{occurring} \\
% 	\text{after cause } \varphi_c \text{ in actual world } \hat{\sigma}}} > \\
% &	 \underbrace{\mathbb{P}^{\text{max}}\Big(\neg \varphi_{\hat{\sigma}'} ^c \U 
% 	 \varphi_{\hat{\sigma}'}^e\Big)}_{\substack{\psi_\CWABS\text{:Maximum probability of cause } \varphi^c  \text{not occurring} \\ 
% 		\text{before effect } \varphi_e \text{ in counterfactual world } \hat{\sigma}'}} ~\land ~\mathbb{P}_{=1} \big(\bigwedge_{p\in W} \G (p_{\sigma} \leftrightarrow 
% 		p_{\sigma'})\big)
% \end{align}
\[
\mathbb{P}^{\text{min}} \Big(\neg \varphi^e_{\hat{\sigma}}~\U~ (\varphi_{\hat{\sigma}}^c \land \mathbb{P}^{\text{min}}_{>0}(\F 
	\varphi_{\hat{\sigma}}^e))\Big) >  \mathbb{P}^{\text{max}}\Big(\neg \varphi_{\hat{\sigma}'} ^c \U 
	 \varphi_{\hat{\sigma}'}^e\Big)
\]
The formula above indicates that the minimum reachability to the effect states in the actual world ($\psi_{\AWABS}$) must be greater than the maximum reachability in the counterfactual world ($\psi_{\CWABS}$). 
In addition, we reason about contingencies in the abstract model using ($\psi_{\SEABS}$).
%
% Here, $\hat{\sigma}$ and $\hat{\sigma}'$ denote state variables ranging over $\hat{\states}$, representing the root states of the actual and counterfactual worlds, respectively, in the abstracted model.
%
% We now present the SMT encoding of the three subformulas: $\psi_\AWABS$, $\psi_\CWABS$, and $\psi_\SEABS$.
\subsubsection{Encoding actual world computations ($\boldsymbol{\psi_{\AWABS}}$).}
\vmcai{We begin by computing $p_{\REABS}$, which corresponds to evaluating $\mathbb{P}^{\min}(\F \varphi^e)$, that is, the minimum probability of reaching the effect in the abstract model. If $\hat{\state} \models \varphi^e$, then $p_{\REABS_{\hat{\state}}} = 1$; if $\hat{\state} \models \neg \varphi^e \land \halt$, then $p_{\REABS_{\hat{\state}}} = 0$. Otherwise:
\[
p_{\REABS_{\hat{\state}}} = \min_{\action \in \Act(\hat{\state})} \sum_{\hat{\state}' \in \hat{\states}} \Pmdp(\hat{\state}, \action, \hat{\state}') \cdot p_{\REABS_{\hat{\state}'}}
\]
}

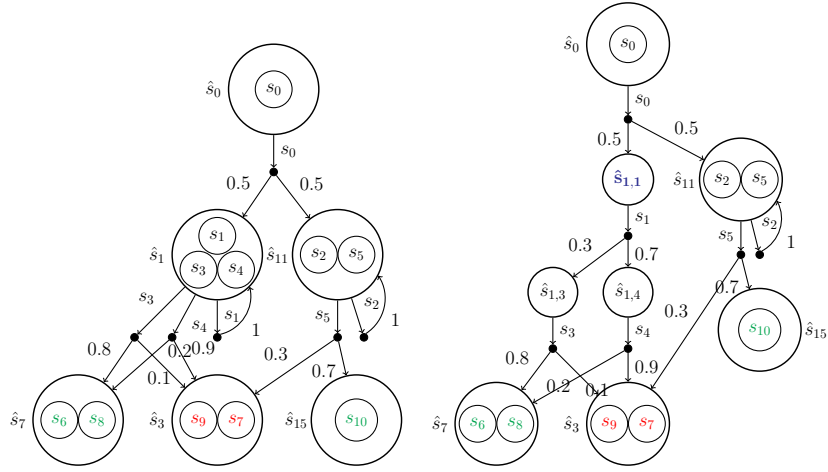
\begin{figure*}[t!]
     \begin{subfigure}[b]{0.45\textwidth}
     \centering
    \scalebox{.5}{\begin{tikzpicture}[->, auto, thick, node distance=2.2cm]
  % First large circle
  \tikzstyle{action}=[circle, fill=black, draw=black,  minimum size=0.2cm,inner sep=0]
  \Large
  
  \node[circle, draw, minimum size=2.4cm, line width=1.1pt] (outer1) {};

  \node[circle, draw, minimum size=0.5cm] at (outer1.center) {$s_0$};
  \node[action]    (s0)          [below of=outer1]      {};

  \node[circle, draw, minimum size=2.4cm, line width=1.1pt, below of=s0, xshift = -1.5cm] (outer2) {};

   \node[action]    (s1)          [below of=outer2]      {};
    \node[action]    (s4)            [left of=s1,xshift=1cm]    {};
    \node[action]    (s3)          [left of=s1,xshift=0cm]      {};
    
   % Three small circles inside the second large one
  \node[circle, draw, minimum size=0.2cm] at ([xshift=-0.5cm, yshift=-0.4cm]outer2.center) {$s_3$};
  \node[circle, draw, minimum size=0.2cm] at ([yshift=0.5cm]outer2.center) {$s_1$};
  \node[circle, draw, minimum size=0.2cm] at ([xshift=0.5cm, yshift=-0.4cm]outer2.center) {$s_4$};

  \node[circle, draw, minimum size=2.4cm, line width=1.1pt, right of=outer2, xshift=1cm] (outer3) {};
\node[circle, draw, minimum size=0.2cm] at ([xshift=-0.5cm]outer3.center) {$s_2$};
  \node[circle, draw, minimum size=0.2cm] at ([xshift=0.5cm]outer3.center) {$s_5$};

     \node[action]    (s5)          [below of=outer3, xshift=0cm]      {};
    \node[action]    (s2)            [right of=s5, xshift=-1.5cm]    {};

    \node[circle, draw, minimum size=2.4cm, line width=1.1pt, below of=s1, xshift=0cm] (outer5) {};
    \node[circle, draw, minimum size=0.2cm] at ([xshift=0.5cm]outer5.center) {\textcolor{red}{$s_7$}};
    \node[circle, draw, minimum size=0.2cm] at ([xshift=-0.5cm]outer5.center) {\textcolor{red}{$s_9$}};

    \node[circle, draw, minimum size=2.4cm, line width=1.1pt, left of=outer5, xshift=-1.5cm] (outer4) {};
    \node[circle, draw, minimum size=0.2cm] at ([xshift=-0.5cm]outer4.center) {\textcolor{Green}{$s_6$}};
  \node[circle, draw, minimum size=0.2cm] at ([xshift=0.5cm]outer4.center) {\textcolor{Green}{$s_8$}};

    \node[circle, draw, minimum size=2.4cm, line width=1.1pt, below of=s5, xshift=0.5cm] (outer6) {};
    
  \node[circle, draw, minimum size=0.2cm] at (outer6.center) {\textcolor{Green}{$s_{10}$}};

  \node[left of=outer1, xshift=0.6cm] {$\hat{s}_0$};
    \node[left of=outer2, xshift=0.6cm] {$\hat{s}_1$};
    \node[left of=outer3, xshift=0.6cm] {$\hat{s}_{11}$};
    \node[left of=outer4, xshift=0.6cm] {$\hat{s}_{7}$};
    \node[left of=outer5, xshift=0.6cm] {$\hat{s}_{3}$};
    \node[left of=outer6, xshift=0.6cm] {$\hat{s}_{15}$};

  \path
			(outer1)
			edge [] node {$s_0$} (s0)
                (s0)
			edge [] node [swap]{0.5} (outer2)
                edge [] node {0.5} (outer3)

                (outer2)
			edge [] node {$s_1$} (s1)
                
                edge [] node {$s_4$} (s4)
                edge [] node [swap]{$s_3$} (s3)

                (s1)
			edge [bend right=45] node [swap]{1} (outer2)

            (s3)
			edge [] node []{0.2} (outer5)
            edge [] node [swap]{0.8} (outer4)

            (s4)
			edge [] node []{0.9} (outer5)
            edge [] node []{0.1} (outer4)

            (outer3)
			edge [] node [swap]{$s_5$} (s5)
            edge [] node []{$s_2$} (s2)

            (s2)
			edge [bend right=45] node [swap]{1} (outer3)

            (s5)
			edge [] node [swap]{0.7} (outer6)
            edge [] node [swap]{0.3} (outer5);

\end{tikzpicture}}
    \caption{Initial abstraction based on predicate set $\predicate$.}
    \label{fig:mdp_example}
\end{subfigure}
%\hspace{.5cm}
     \begin{subfigure}[b]{0.45\textwidth}        
     \centering
        \scalebox{0.5}{
        \begin{tikzpicture}[->, auto, thick, node distance=2cm]
  % First large circle
  \tikzstyle{action}=[circle, fill=black, draw=black,  minimum size=0.2cm,inner sep=0]
  \Large
  
  \node[circle, draw, minimum size=2.2cm, line width=1.1pt] (outer1) {};

  \node[circle, draw, minimum size=0.5cm] at (outer1.center) {$s_0$};
  \node[action]    (s0)          [below of=outer1]      {};

  \node[circle, draw, minimum size=0.5cm, yshift=0.4cm,line width=1.1pt, below of=s0, xshift = 0cm] (s11) {\textcolor{Blue}{$\mathbf{\hat{s}_{1,1}}$}};

  \node[action]    (s1)          [yshift=0.5cm,below of=s11]      {};
  \node[circle, draw, minimum size=0.5cm, line width=1.1pt, yshift=0.5cm,below of=s1, xshift = 0cm] (s14) {$\hat{s}_{1,4}$};
  \node[circle, draw, minimum size=0.5cm, line width=1.1pt, left of=s14, xshift = 0cm] (s13) {$\hat{s}_{1,3}$};

    \node[action]    (s4)            [yshift=0.5cm,below of=s14]    {};
    \node[action]    (s3)          [yshift=0.5cm,below of=s13,xshift=0cm]      {};

  \node[circle, draw, minimum size=2.2cm, line width=1.1pt, right of=s11, xshift=1cm] (outer3) {};
\node[circle, draw, minimum size=0.2cm] at ([xshift=-0.5cm]outer3.center) {$s_2$};
  \node[circle, draw, minimum size=0.2cm] at ([xshift=0.5cm]outer3.center) {$s_5$};

     \node[action]    (s5)          [below of=outer3, xshift=0cm]      {};
    \node[action]    (s2)            [right of=s5, xshift=-1.5cm]    {};

    \node[circle, draw, minimum size=2.2cm, line width=1.1pt, below of=s4, xshift=0cm] (outer5) {};
    \node[circle, draw, minimum size=0.2cm] at ([xshift=0.5cm]outer5.center) {\textcolor{red}{$s_7$}};
    \node[circle, draw, minimum size=0.2cm] at ([xshift=-0.5cm]outer5.center) {\textcolor{red}{$s_9$}};

    \node[circle, draw, minimum size=2.2cm, line width=1.1pt, left of=outer5, xshift=-1.5cm] (outer4) {};
    \node[circle, draw, minimum size=0.2cm] at ([xshift=-0.5cm]outer4.center) {\textcolor{Green}{$s_6$}};
  \node[circle, draw, minimum size=0.2cm] at ([xshift=0.5cm]outer4.center) {\textcolor{Green}{$s_8$}};

    \node[circle, draw, minimum size=2.2cm, line width=1.1pt, below of=s5, xshift=0.5cm] (outer6) {};
    
  \node[circle, draw, minimum size=0.2cm] at (outer6.center) {\textcolor{Green}{$s_{10}$}};

  \node[left of=outer1, xshift=0.5cm] {$\hat{s}_0$};
    \node[left of=outer3, xshift=0.5cm] {$\hat{s}_{11}$};
    \node[left of=outer4, xshift=0.5cm] {$\hat{s}_{7}$};
    \node[left of=outer5, xshift=0.5cm] {$\hat{s}_{3}$};
    \node[right of=outer6, xshift=-0.5cm] {$\hat{s}_{15}$};

 \path
			(outer1)
			edge [] node {$s_0$} (s0)
                (s0)
			edge [] node [swap]{0.5} (s11)
                edge [] node {0.5} (outer3)

                (s11)
			edge [] node {$s_1$} (s1)

                (s1)
			edge [] node {0.7} (s14)
            edge [] node [swap]{0.3} (s13)

            (s13)
		edge [] node {$s_3$} (s3)

        (s14)
		edge [] node {$s_4$} (s4)

            (s3)
			edge [] node [swap]{0.2} (outer5)
            edge [] node [swap]{0.8} (outer4)

            (s4)
			edge [] node []{0.9} (outer5)
            edge [] node {0.1} (outer4)

            (outer3)
			edge [] node [swap]{$s_5$} (s5)
            edge [] node []{$s_2$} (s2)

            (s2)
			edge [bend right=45] node [swap]{1} (outer3)

            (s5)
			edge [] node [swap]{0.7} (outer6)
            edge [] node [swap]{0.3} (outer5);

\end{tikzpicture}}
        \caption{Abstract model after one step of refinement.}
        \label{fig:refine}
        \end{subfigure}
\caption{Concrete and Abstract model.}
    \end{figure*}

\vmcai{Next, we compute $p_{\AWABS}$, which corresponds to evaluating $\psi_{\AWABS}$. We set $p_{\AWABS_{\hat{\state}}} = 1$ if $(\hat{\state} \models \varphi^c) \land (p_{\REABS_{\hat{\state}}} > 0)$. If $(\hat{\state} \models \neg \varphi^c \land (\halt \lor \varphi^e)) \lor (p_{\REABS_{\hat{\state}}} = 0)$, we set $p_{\AWABS_{\hat{\state}}} = 0$. Otherwise:
\[
p_{\AWABS_{\hat{\state}}} = \min_{\action \in \Act(\hat{\state})} \sum_{\hat{\state}' \in \hat{\states} \setminus \varphi^e} \Pmdp(\hat{\state}, \action, \hat{\state}') \cdot p_{\mathsf{\AWABS}_{\hat{\state}'}}
\]}

\subsubsection{Encoding counterfactual world computations ($\boldsymbol{\psi_{\CWABS}}$).} We introduce a variable $p_{\CWABS_{\hat{\state}}}$, such that if $\hat{\state} \models \varphi^e$, then $p_{\CWABS_{\hat{\state}}} =1$. 
If $\hat{\state} \models \neg \varphi^e \land (\halt \lor \varphi^c)$, then $p_{\CWABS_{\hat{\state}}}=0$. Otherwise:
% $$
% p_{\CWABS_{\hat{\state}}}^{\min}  = \min_{\action \in \Act(\hat{\state})} \sum_{\hat{\state}' \in \hat{\states} \setminus \varphi^c} \Pmdp(\hat{\state}, \action, \hat{\state}') \cdot p_{\CWABS_{\hat{\state}'}}^{\min} 
% $$
\[
p_{\CWABS_{\hat{\state}}}  = \max_{\action \in \Act(\hat{\state})} \sum_{\hat{\state}' \in \hat{\states} \setminus \varphi^c} \Pmdp(\hat{\state}, \action, \hat{\state}') \cdot p_{\CWABS_{\hat{\state}'}}
\]
\subsubsection{Encoding equivalence of contingencies ($\boldsymbol{\psi_{\SEABS}}$).}
To reason about $\psi_{\SEABS}$, we consider two cases. 
If we use the enumerating subgraph strategy, then $\psi_{\SEABS}$ is trivially satisfied and does not need to be encoded. 
However, if we use a $W$-preserving abstraction, we must encode $\psi_{\SEABS}$ explicitly. 
%
% This can be done by computing the minimum reachability probabilities for $\psi_{\SE}$, similar to the approach used for the previous two subformulas discussed in this section.
\vmcai{To encode $\psi_{\SEABS}$, we modify the $\SE$ defined in \Cref{sec:smt} and introduce $\SEABS$, defined as:
\[ 
\SEABS_{\hat{\state}, \hat{\state}'} \triangleq \forall \scheduler,\scheduler' \in \Schedulers ~.~  \forall \pi \in \fPaths{\hat{\state},\haltstate}{\hat{\dtmc}^\scheduler} ~ . ~ \forall \pi' \in \fPaths{\hat{\state}',\haltstate}{\hat{\dtmc}^{\scheduler'}} ~. ~ \ST_{\pi,\pi'} 
\]
where $\Schedulers$ denotes the finite set of memoryless schedulers available in the abstract model $\hat{\dtmc}$.
}

The rest of the SMT formulation, including the uninterpreted function representing the actual world and the SMT variables $\varphi^c$ and $\varphi^e$, remains consistent with the details presented in Section~\ref{sec:smt}. 

To proceed with abstraction-refinement algorithm, we search for a cause using the SMT-based approach described in this section. 
If we find a cause, it is returned as the probabilistic actual cause of the effect. 
Otherwise, the absence of a cause suggests that the current abstract model is too coarse to capture the underlying causality and must therefore be refined.

\begin{example}
\label{ex:abs-smt}
Continuing with \Cref{ex:pred}, consider the abstract model shown in~\Cref{fig:mdp_example} and the formula $\varphi^\fail$. Our goal is to solve the decision problem of identifying $\varphi^c$. 
\begin{itemize}
    \item First, it is clear that states $\hat{s}_7$ and $\hat{s}_{15}$ are not candidates, since their minimum and maximum reachability probabilities to effect are both 0. Similarly, $\hat{s}_3$ is excluded because it is labeled with $\varphi^\fail$. State $\hat{s}_0$ is also not a valid candidate, since it is always included in both actual and counterfactual computation trees.
    \item Consider $\hat{s}_1$ as a candidate. We find that the minimum probability of reaching $\hat{s}_3$ through $\hat{s}_1$ is $\mathbb{P}^{\text{min}} (\neg \varphi^e_{\hat{\sigma}}~\U~ (\varphi_{\hat{\sigma}}^c \land \mathbb{P}^{\text{min}}_{>0}(\F \varphi_{\hat{\sigma}}^e))) = 0.5 \times 0.2 = 0.1$. On the other hand, in the counterfactual world, the maximum probability is $\mathbb{P}^{\text{max}}=(\neg \varphi_{\hat{\sigma}'} ^c \U \varphi_{\hat{\sigma}'}^e) =0.5 \times 0.3 = 0.15$. Since $0.1 \not> 0.15$, $\hat{s}_1$ does not satisfy the condition.
    \item Next, consider $\hat{s}_{11}$. Using the same reasoning, we find that the minimum probability of reaching the effect in the actual world is $0.15$, while the maximum in the counterfactual world is $0.45$. Again, $0.15 \not> 0.45$, so this state also fails the condition.
\end{itemize}
Consequently, none of the abstract states satisfy the specification. Since we couldn't find a valid cause, we conclude that the current abstraction is too coarse and proceed with refinement, described next.\qed
\end{example}

\subsection{Refinement}
\label{sec:ref}

As discussed in \Cref{sec:smt-abs}, if a cause cannot be found, then the abstract MDP is possibly too coarse and must be refined to a lower level of abstraction. 
In the refinement process, the coarser model $\hat{\dtmc}_i {=} \big(\hat{\states}_i, \mathcal{P}_i, \AP, \hat{L}_i, \Act_i\big)$ is refined to a less coarse model $\hat{\dtmc}_{i+1}$ by identifying the abstract state $\hat{\state}_{\Delta}$. \vmcai{We employ two heuristic methods to identify $\hat{\state}_{\Delta}$:}
\begin{enumerate}

\item \textbf{Number of Available Actions}: The state with the maximum number of available actions in $\hat{\dtmc}_i$.
   \[
\hat{\state}_{{\Delta}_i} = \underset{\hat{s}\in \hat{\states}_i}{\mathrm{arg max}} ~~ \Big| \Act(\hat{s}) \Big|
\]
    \item \textbf{Range of Reachability:} The state with the maximum range between its minimum ($\mathbb{P}^{\min}(\F \varphi^e)$) and maximum ($\mathbb{P}^{\max}(\F \varphi^e)$) reachability to the effect states in the $\hat{\dtmc}_i$.
    \[
\hat{\state}_{{\Delta}_i} = \underset{\hat{s}\in \hat{\states}_i}{\mathrm{arg max}} ~~ \Big(\mathbb{P}^{\max}(\hat{s} \models \F \varphi^e) - \mathbb{P}^{\min}(\hat{s} \models \F \varphi^e)\Big)
\]

\end{enumerate}

Once $\hat{\state}_{{\Delta}_i}$ is identified, we refine it by splitting the underlying concrete states it represents, thereby constructing a less coarse abstract model $\hat{\dtmc}_{i+1}$ for the next iteration. 
This refinement step induces a new MDP \linebreak $\dtmc_{i+1} {=} \big(\hat{\states}_{i+1}, \mathcal{P}_{i+1}, \AP, \hat{L}_{i+1}, \Act_{i+1}\big)$, where $\hat{\states}_{i+1} = (\states_i \setminus \hat{s}_{\Delta_i}) ~\cup ~ \gamma(\hat{s}_{\Delta_i})$, $\Act_{i+1}(\hat{s}) $ $= \{\, \action \in \states \mid \sum_{\hat{s}' \in \hat{\states}_{i+1}} \mathcal{P}_{i+1}(\hat{s}, \action, \hat{s}') = 1 \,\}$, $\mathcal{P}_{i+1}(\hat{s}, \action, \hat{s}') = \sum_{s'\in \gamma(\hat{s}')} \P(\action, s')$, and $\hat{L}_{i+1}(\hat{\state}) = \bigcup_{s \in \gamma(\hat{s})} L(s)$.
% \begin{itemize}
%     \item  $\hat{\states}_{i+1}, = \states_i \bigcup \gamma(\hat{s}_{\Delta})$
%     \item $\Act_{i+1}(\hat{s}) = \{\, \action \in \states \mid \sum_{\hat{s}' \in \hat{\states}_{i+1}} \mathcal{P}_{i+1}(\hat{s}, \action, \hat{s}') = 1 \,\}$;

%     \item $\mathcal{P}_{i+1}(\hat{s}, \action, \hat{s}') = \sum_{s'\in \gamma(\hat{s}')}\Pmdp(\action, s')$, and 

%     \item $\hat{L}_{i+1}(\hat{\state}) = \bigcup_{s \in \gamma(\hat{s})} L(s)$.
% \end{itemize}
%
%Implementation details on the splitting strategies are provided in \Cref{sec:exp}.

\begin{example}
    Continuing from \Cref{ex:abs-smt}, we observe that the initial abstract model shown in~\Cref{fig:mdp_example} is too coarse and requires refinement. 
    \vmcai{In this model, $\hat{\state}_{\Delta}$ corresponds to $\hat{s}_1$ according to both heuristics, as it exhibits the widest reachability range (i.e., [0.2, 0.9]) and possesses the highest number of available actions (i.e., 3).}
    To refine the abstraction, we split $\hat{s}_1$ into three more precise abstract states: $\hat{s}_{1,1}$, $\hat{s}_{1,3}$, and $\hat{s}_{1,4}$. The refined abstract model is shown in~\Cref{fig:refine}. 
    In this refined model, if we consider $\hat{s}_{1,1}$ as a candidate, we compute the minimum probability of reaching $\hat{s}_3$ through $\hat{s}_{1,1}$ in the actual world as, $\mathbb{P}^{\text{min}} (\neg \varphi^e_{\hat{\sigma}}~\U~ (\varphi_{\hat{\sigma}}^c \land \mathbb{P}^{\text{min}}_{>0}(\F \varphi_{\hat{\sigma}}^e))) = 0.5 \times 0.7 \times 0.9 + 0.5 \times 0.3 \times 0.2 = 0.345$. 
    On the other hand, in the counterfactual world, the maximum probability of reaching the effect is $\mathbb{P}^{\text{max}}=(\neg \varphi_{\hat{\sigma}'} ^c \U \varphi_{\hat{\sigma}'}^e) =  0.5 \times 0.3 = 0.15$.
    Since $0.345 > 0.15$, $\hat{s}_{1,1}$ satisfies the specifications.~\qed
\end{example}

%\subsection{Correctness}
%In this section, we formally prove the soundness of Algorithm~\ref{alg:absref}\footnote{Proof can be found in \Cref{app:proof}}.
\begin{theorem}
\label{th:sound}
Let $\dtmc$ be a concrete causal model and $\varphi_c$ and $\varphi_e$ be two 
	predicates.
	If $\varphi_c$ is an actual cause of $\varphi_e$ identified by our abstraction-refinement technique, then $\varphi_c$ is an actual cause of $\varphi_e$ in $\dtmc$.
\end{theorem}
% \begin{proof}
%     \begin{itemize}
%         \item minimum reach is greater than 0 so at worst case there is a reachibilty to effect.
%         \item min reach of actual  is greater than max reach of counterfactual in a dtmc where the W are same. 
%     \end{itemize}
% \end{proof}

\section{Experimental Evaluation}
\label{sec:exp}

In this section, we evaluate our approach by applying the techniques discussed in \Cref{sec:smt,sec:abs-ref} to three case studies: (1) the Mountain Car and (2) Lunar Lander environments from {\sf \small OpenAI Gym}~\cite{openaigym}, and (3) an F-16 autopilot simulator~\cite{ARCH18:Verification_Challenges_in_F_16} that employs an MPC controller.
\footnote{All implementation artifacts are  available at 
\texttt{\url{https://github.com/rogaleke/Prob-HP}}}

\subsection{Implementation}
\label{sec:impl}
\subsubsection{DTMC generation.}
\label{sec:dtmc-gen}
The environments used in our case studies are originally deterministic. 
To introduce stochastic behavior (e.g., noise), we synthesize DTMCs for each environment.
To generate a DTMC, we begin by constructing random probability vectors whose components sum to one. 
For a state with $k$ outgoing transitions, we sample $k$ values $u_i \sim U(0,1)$ and normalize them as $p_i = \frac{u_i}{\sum_{j=1}^{k} u_j}$. 
The value of $k$ is randomly chosen, allowing control over the degree of branching in the DTMC.
Each distribution defines transitions to successor states, generated by adding noise to the original successor states.
We continue this process up to a fixed number of steps to control the overall size of the resulting DTMC.

\subsubsection{Algorithm.}
We used the Python programming language along with the Python API of the \textsf{Z3} SMT solver~\cite{dmb08}. 
\vmcai{In general, \textsf{Z3} is incomplete for non-linear arithmetic; however, it successfully handles all our benchmarks, as the considered DTMC models are bounded and acyclic.}
Our implementation involves two key hyperparameters: the initial set of predicates and the splitting strategy for refining an abstract state $\hat{\state}_{\Delta}$. 
\camera{
In order to control the splitting strategy, we introduced a tunable hyperparameter $\beta \in (0, 1]$ which controls the fraction of an abstract state to be concretized when refining the abstract state. 
This fraction determines the trade-off between the number of iterations required to identify a cause and the size of the model in the subsequent iteration.
}
It is important to note that the performance of the abstraction-refinement technique is highly sensitive to these hyperparameters. 
For example, an unsuitable initial predicate set may result in a model that is too coarse, requiring several rounds of refinement to discover the cause. 
On the other hand, using a finer abstraction generates a large number of abstract states, thereby diminishing the performance advantages of abstraction and resulting in solving times comparable to those of the concrete model. 
\vmcai{Moreover, a poor choice of predicates may result in abstract states that conflate concrete states where the effect holds with those where it does not. 
This ambiguity can lead to difficulties in computing reachability probabilities, as it becomes unclear whether such an abstract state should be considered as satisfying the effect.
Therefore, it is crucial to select predicates that distinguish states based on key properties, including the effect.}
Further details on splitting strategies are provided in \Cref{app:hyp}.

\subsection{Experimental Settings}

All of our experiments were conducted on a single core of the Intel i9-12900K CPU, which features a 
16-core architecture and operates @5.2GHz. 

%
%We have repeated each experiment 1000 times to gain at least 95\% confidence interval.

%\vspace{-2mm}
\subsection{Case Study 1: Mountain Car}
\label{sec:car}

Our first case study extends the running example. 
As illustrated in Fig.~\ref{fig:car_M}, the car begins in a valley between two mountains with the objective of reaching the peak of the right hand mountain before a specified time-bound.
%
% The action $\act$ is selected by a pre-trained reinforcement learning policy implemented as a deep neural network. 
%
The action controller in the Mountain Car scenario uses a Deep Q-Network (DQN) based on the model provided in~\cite{stable-baselines3}, and is trained using the hyperparameters from~\cite{rl-zoo3}.
The objective of this case study is to discover the cause of failure, defined as the car failing to reach the target position within a fixed number of steps.
To construct DTMCs, we execute the pre-trained controller under multiple initial valuations, following the procedure detailed in \Cref{sec:dtmc-gen}.

\subsection{Case Study 2: Lunar Lander}
\label{sec:lunar}

In this case study, a lunar lander begins at a certain altitude with the goal of landing on a designated landing pad. 
The Lunar Lander environment includes eight variables (including $x$ and $y$ coordinates, linear and angular velocities, angle, etc.).
Failure in this case study is defined as failing to land safely on the landing pad within a designated number of steps.
The action controller is implemented using Proximal Policy Optimization (PPO), following the model architecture from~\cite{stable-baselines3} and trained with hyperparameters from~\cite{rl-zoo3}. 
The controller operates over four discrete actions available in the environment, defined as $\act = \{0, 1, 2, 3\}$, where $0$ corresponds to doing nothing, $1$ fires the left orientation engine, $2$ fires the main engine, and $3$ fires the right orientation engine.
We construct DTMCs of varying sizes by executing the pre-trained controller under multiple initial valuations and varying parameters such as the number of simulation steps, while introducing uncertainty through noise injection. 

\subsection{Case Study 3: F-16 Autopilot MPC Controller~\cite{ARCH18:Verification_Challenges_in_F_16}}
\label{sec:f16}

This benchmark models the outer-loop controller of the F-16 fighter jet.
We examine two scenarios: the first focuses on reaching a specified speed, and the second on achieving a target altitude.

\subsubsection{First Scenario}
\label{sec:f16_1}
The first scenario investigates how the engine responds under control inputs. 
The simulation models two variables: airspeed $\speed$ and engine power lag state $\pow$, with all remaining variables held constant. 
Control is applied to adjust $\speed$ toward a setpoint value using the throttle input $\throt$.
The scenario starts with initial values for $\speed$ and $\alt$ and aims to verify whether the system can reach the desired $\speed$ setpoint within a specified time.
In this scenario, we investigate the causes of failure, which are categorized as either failing to reach the target $\speed$ within the designated time or violating the safety constraints of the aircraft.

\begin{table}[b!]
    \centering
    \caption{\textbf{Mountain Car}: Comparing Abstraction-Refinement (Two Heuristics) and Concrete SMT Solving, Averaged Across 10 Independent Runs.}
    \begin{tabular}{c || c || c ||c c |c|| c c |c}
       \multicolumn{3}{c}{} & \multicolumn{3}{c}{\textbf{Abs (Act Based)}} & \multicolumn{3}{c}{\textbf{Abs (Range Based)}} \\
      \cmidrule(lr){4-6} \cmidrule(lr){7-9}
       \textbf{Case} &  $|\states|$ &  \textbf{Conc(s)}   & Abs-Ref(s) & SMT(s) & \textbf{Tot(s)} & Abs-Ref(s) & SMT(s) & \textbf{Tot(s)} \\
       \hline
      \multirow{9}{*}{\rotatebox[origin=c]{90}{ \textbf{Mountain Car}}}
      & 203       & 1.41 & 0.02 & 0.64  & \underline{0.67} &   0.02 & 0.31  & \textbf{0.33}  \\
& 359       & 3.22 &   0.02 & 2.52  & \underline{2.53}    & 0.02 & 1.82  & \textbf{1.84}  \\
& 1221      & 14.34 &     0.06 & 0.88  & \textbf{0.94} & 0.06 & 0.87  & \textbf{0.94}  \\
& 1943      & 4.97 &      0.06 & 2.08  & \textbf{2.14}  & 0.06 & 2.12  & \underline{2.18}  \\
& 2261      & 9.85 &      0.09 & 4.73  & \textbf{4.82}  & 0.09 & 4.93  & \underline{5.02}  \\
& 2603      & 12.41 &    0.09 & 6.38  & \textbf{6.46}  & 0.08 & 6.46  & \underline{6.54}  \\
& 2888      & 1.78   &    0.08 & 1.15  & \textbf{1.23}  &  0.08 & 1.15  & \textbf{1.23}  \\
& 4558      & 40.19 &    0.19 & 12.69 & \underline{12.89} & 0.19 & 12.59 & \textbf{12.79} \\
% & 16648     & \textbf{0.71}  &    0.38 & 0.61  & 0.99  & 0.37 & 0.61  & \underline{0.98}  \\
& 27547     & 18.37  &    0.51 & 0.70   & \textbf{1.21} & 0.51 & 0.71  & \underline{1.22 }
    \end{tabular}
    \label{tab:mcar-results}
\end{table}

\subsubsection{Second Scenario}
\label{sec:f16_2}
In this scenario, the aircraft is expected to reach a specific target altitude $\alt$, and it is flying without any roll or yaw.
The inputs to the system are the engine throttle $\throt$ and the elevator $\elev$. The model includes seven state variables: airspeed $\speed$, angle of attack $\aoa$, pitch angle $\pit$, pitch rate $\pitrate$, altitude $\alt$, engine power lag $\pow$, and upward acceleration (G-force) $\gforce$.
In this scenario, failures are defined as either not reaching the required altitude $\alt$ within the given time frame or violating safety parameters, such as limits on G-force and angle of attack.

\begin{table}[b!]
    \centering
    
        \caption{\textbf{Lunar Lander}: Comparing Abstraction-Refinement (Two Heuristics) and Concrete SMT Solving, Averaged Across 10 Independent Runs.}
    \begin{tabular}{c || c || c ||c c |c|| c c |c}
       \multicolumn{3}{c}{} & \multicolumn{3}{c}{\textbf{Abs (Act Based)}} & \multicolumn{3}{c}{\textbf{Abs (Range Based)}} \\
      \cmidrule(lr){4-6} \cmidrule(lr){7-9}
       \textbf{Case} &  $|\states|$ &  \textbf{Conc(s)}   & Abs-Ref(s) & SMT(s) & \textbf{Tot(s)} & Abs-Ref(s) & SMT(s) & \textbf{Tot(s)} \\
       \hline
      \multirow{10}{*}{\rotatebox[origin=c]{90}{ \textbf{Lunar Lander}}}
        & 115 & \textbf{0.07} & 0.01 & 0.07 & 0.08 & 0.01 & 0.07 & 0.08  \\ 
         & 192 & \textbf{0.09} & 0.01 & 0.19 & 0.20 & 0.01 & 0.20 & 0.21  \\ 
         & 1381 & 0.59 & 0.04 & 0.32 & \underline{0.36} & 0.04 & 0.29 & \textbf{0.33}  \\ 
         & 2458 & 3.56 & 0.07 & 0.62 & \underline{0.68} & 0.06 & 0.53 & \textbf{0.59}  \\ 
         & 3431 & 9.71 & 0.09 & 1.33 & \textbf{1.41} & 0.09 & 2.65 & \underline{2.74}  \\ 
         & 5670 & \underline{6.63} & 0.32 & 18.31 & 18.63 & 0.57 & 3.44 & \textbf{4.01}  \\ 
         & 9282 & 51.36 & 0.23 & 1.16 & \textbf{1.38} & 0.22 & 1.32 & \underline{1.54}  \\ 
         & 11653 & 24.96 & 0.30 & 0.43 & \underline{0.74} & 0.30 & 0.43 & \textbf{0.73}  \\ 
         & 14200 & 37.20 & 0.35 & 3.05 & \textbf{3.39} & 0.34 & 4.73 & \underline{5.07}  \\ 
         & 47915 & 39.15 & 1.20 & 9.89 & \underline{11.09} & 1.20 & 8.73 & \textbf{9.92} \\ 
    \end{tabular}
    \label{tab:lunar-results}
\end{table}
\subsection{Performance Analysis}
\label{sec:prefanalysis}
\vmcai{\Cref{tab:mcar-results,tab:lunar-results} summarize the results for the Mountain Car and Lunar Lander environments, and \Cref{tab:f161-results,tab:f162-results} present the outcomes for the two F-16 scenarios. 
Bold numbers indicate the best results, and underlined numbers denote the second-best results.
As shown in the tables, the abstraction-refinement algorithm (employing both refinement heuristics) achieves significantly better performance in discovering probabilistic actual causes compared to concrete SMT solving especially in larger DTMCs.
In most experiments, both refinement heuristics achieve comparable performance, with the range-based heuristic performing slightly better in the Mountain Car, Lunar Lander, and second F-16 scenario, while the action-based heuristic shows an advantage in the first F-16 scenario.
In addition, in some smaller DTMCs, the overhead of constructing abstract models and performing refinement steps may outweigh the benefits, making direct SMT solving on the concrete model more efficient for identifying causes.}
However, for larger DTMCs especially in the F-16 first scenario and second scenario, the performance gains from abstraction-refinement are substantial.

\begin{table}[t!]
    \centering
    
    \caption{\textbf{F-16 1st Scenario}: Comparing Abstraction-Refinement (Two Heuristics) and Concrete SMT Solving, Averaged Across 10 Independent Runs.}
    \begin{tabular}{c || c || c ||c c |c|| c c |c}
       \multicolumn{3}{c}{} & \multicolumn{3}{c}{\textbf{Abs (Act Based)}} & \multicolumn{3}{c}{\textbf{Abs (Range Based)}} \\
      \cmidrule(lr){4-6} \cmidrule(lr){7-9}
       \textbf{Case} &  $|\states|$ &  \textbf{Conc(s)}   & Abs-Ref(s) & SMT(s) & \textbf{Tot(s)} & Abs-Ref(s) & SMT(s) & \textbf{Tot(s)} \\
       \hline
      \multirow{10}{*}{\rotatebox[origin=c]{90}{ \textbf{F-16 1st scenario}}}
        & 84 & 1.47  & 0.01 & 0.10 & \textbf{0.11}  & 0.04  & 0.40 & \underline{0.44}  \\ 
         & 130 & 1.64  & 0.01 & 0.13 & \textbf{0.14}  & 0.02  & 0.31 & \underline{0.32}  \\ 
         & 371 & 18.10  & 0.05 & 4.31 & \underline{4.36}  & 0.03  & 1.99 & \textbf{2.02}  \\ 
         & 401 & 15.24  & 0.02 & 0.53 & \textbf{0.55}  & 0.03  & 0.83 & \underline{0.87}  \\ 
         & 599 & 32.16  & 0.08 & 8.30 & \underline{8.38}  & 0.04  & 3.84 & \textbf{3.88}  \\ 
         & 630 & 50.64  & 0.05 & 3.48 & \textbf{3.52}  & 0.08  & 4.12 & \underline{4.20}  \\ 
         & 2511 & 883.68  & 0.19 & 41.15 & \textbf{41.33}  & 0.38  & 72.47 & \underline{72.85}  \\ 
         & 3405 & 1287.35  & 0.30 & 137.66 & \textbf{137.95}  & 0.21  & 145.86 & \underline{146.07}  \\ 
         & 4175 & 1028.07  & 0.19 & 33.90 & \textbf{34.09}  & 0.37  & 56.58 & \underline{56.96}  \\ 
         & 13842 & 9582.02 & 1.00 & 633.01 & \underline{634.01} & 0.50 & 395.62 & \textbf{396.12} 
    \end{tabular}
    \label{tab:f161-results}
\end{table}

Additionally, we observe a meaningful correlation between execution time and the size of the DTMC. 
Notably, performance is also influenced by the depth-width ratio of the DTMC structure. 
In our experiments, deeper DTMCs with lower branching factors (as observed in F-16 scenarios) make cause discovery more challenging for both the abstraction-refinement and concrete SMT approaches. 
For instance, comparing the F-16 first scenario with 13k states in \Cref{tab:f161-results} to the Lunar Lander with 14k states in \Cref{tab:lunar-results} reveals a significant performance difference. 
This discrepancy is due to differences in the DTMC structures: the F-16 model is deeper with less branching, while the Lunar Lander DTMC is shallower with more branching. 
Consequently, the execution time for causal discovery depends on both the number of states and the structure of the DTMC.

\begin{table}[b!]
    \centering
    
        \caption{\textbf{F-16 2nd Scenario}: Comparing Abstraction-Refinement (Two Heuristics) and Concrete SMT Solving, Averaged Across 10 Independent Runs.}
    \begin{tabular}{c || c || c ||c c |c|| c c |c}
       \multicolumn{3}{c}{} & \multicolumn{3}{c}{\textbf{Abs (Act Based)}} & \multicolumn{3}{c}{\textbf{Abs (Range Based)}} \\
      \cmidrule(lr){4-6} \cmidrule(lr){7-9}
       \textbf{Case} &  $|\states|$ &  \textbf{Conc(s)}   & Abs-Ref(s) & SMT(s) & \textbf{Tot(s)} & Abs-Ref(s) & SMT(s) & \textbf{Tot(s)} \\
       \hline
      \multirow{10}{*}{\rotatebox[origin=c]{90}{ \textbf{F-16 2nd scenario}}}
         & 101 & \textbf{1.03}  & 0.06 & 4.62 & 4.67  & 0.03 & 1.32 & \underline{1.35}  \\ 
         & 187 & \textbf{3.57}  & 0.04 & 6.71 & 6.76  & 0.04 & 3.72 & \underline{3.76}  \\ 
         % & 299 & 7.86  & 0.03 & 1.53 & \underline{1.56}  & 0.02 & 1.50 & \textbf{1.52}  \\ 
         & 403 & \underline{13.19}  & 0.05 & 39.59 & 39.63  & 0.03 & 6.13 & \textbf{6.15}  \\ 
         & 417 & 15.54  & 0.03 & 3.70 & \textbf{3.73}  & 0.03 & 5.90 & \underline{5.93}  \\ 
         & 425 & 41.99  & 0.03 & 2.08 & \textbf{2.11}  & 0.04 & 2.39 & \underline{2.43}  \\ 
         & 621 & 33.83  & 0.04 & 22.24 & \underline{22.28}  & 0.04 & 19.10 & \textbf{19.14}  \\ 
         & 879 & 195.42  & 0.05 & 7.87 & \textbf{7.92}  & 0.07 & 9.97 & \underline{10.04}  \\ 
         & 1285 & 153.02  & 0.09 & 141.37 & \underline{141.45}  & 0.17 & 100.49 & \textbf{100.66}  \\ 
         & 1925 & 427.00  & 0.10 & 88.66 & \underline{88.76}  & 0.10 & 29.66 & \textbf{29.75}  \\ 
         & 6535 & 7463.79 & 0.17 & 490.09 & \underline{490.26} & 0.16 & 319.36 & \textbf{319.52} 
    \end{tabular}
    \label{tab:f162-results}
\end{table}

\subsection{Causality Analysis}

\begin{figure*}[t!]
    \centering
    \begin{minipage}{0.45\linewidth}
        \centering
        \resizebox{\linewidth}{!}{\input{images/f16CA}}
        \caption{The F-16 actual world computation tree.} 
        \label{fig:f16-CA}
    \end{minipage}
    \hfill
    \begin{minipage}{0.54\linewidth}
            \centering
            \resizebox{\linewidth}{!}{\input{images/f16CAcount}}
            \caption{The F-16 counterfactual world computation tree.} 
            \label{fig:f16-CAcount}
    \end{minipage}

\end{figure*}

In this section, we apply the causal discovery method to the DTMCs generated from an F-16 simulation scenario by counterfactual computation trees where the probability of failure is lower than in the actual world.
Causal analysis for the Lunar Lander case study is provided in \Cref{app:extra-causal}.

\Cref{fig:f16-CA} illustrates a scenario from the F-16 Autopilot Simulation where the aircraft starts at an altitude of 1000ft and is expected to reach $1400 \pm \epsilon$ ft within a specific time frame.
Failure occurs either when the aircraft does not reach the target altitude within the time limit or violates one of the safety constraints such as G-force ($\gforce$) or angle of attack ($\aoa$).
Our SMT solving technique identifies $\varphi^c \triangleq (\alt = 1293 \land \speed = 610\land \aoa = -1.1\land \pit = 13\land \gforce = -1.4)$ and the computation tree observable in \Cref{fig:f16-CA} is the actual world, as a result, \Cref{fig:f16-CAcount} is the counterfactual world. 
In this simulation, the aircraft starts to gain altitude by adjusting the elevators and throttles. 
However, due to noise (e.g., turbulence), it may reach one of the following three states: $s_1 \triangleq (\alt = 1087 \land \speed = 601 \land \aoa = 10.3 \land \pit = 24 \land \gforce = 2.52 )$ with probability 0.32, $s_2 \triangleq  (\alt = 1132 \land \speed = 665 \land \aoa = 9.7 \land \pit = 29 \land \gforce = 2.76 )$ with probability 0.22, and $s_3 \triangleq (\alt = 1021 \land \speed = 619 \land \aoa = 9.1 \land \pit = 28 \land \gforce = 2.72)$ with probability 0.47.

When the aircraft reaches state $s_1$, it continues to gain altitude. 
At this point, due to noise, the aircraft may enter one of four possible states.
Among these, one state is identified as a probabilistic actual cause of failure $ \varphi^c \triangleq (\alt = 1293$, $\speed = 610$, $\aoa = -1.1$, $\pit = 13$, and $\gforce = -1.4)$, where the probability of reaching failure (specifically due to a violation of the G-force constraint) is 1. 
In contrast, the counterfactual world produces three alternative successor states. 
One of them, $s_4 \triangleq (\alt = 1349 \land \speed = 556 \land \aoa = -1.1 \land \pit = 12 \land \gforce = -1.3)$ with a probability of 0.35, successfully reaches the target altitude without violating safety constraints. 
Another, $s_5 \triangleq (\alt = 1167 \land \speed = 568 \land \aoa = -1.1 \land \pit = 12 \land \gforce = -1.38)$, has a probability of zero for reaching the effect (i.e., failure).
The third, $s_6 \triangleq (\alt = 1241 \land \speed = 552 \land \aoa = -1.1 \land \pit = 12 \land \gforce = -1.28)$, leads to failure with a probability of 0.19. 

A closer examination of the dynamics reveals that $s_6$ and the identified causal state $\varphi^c$ are similar in terms of system variables. 
In both $s_6$ and $\varphi^c$, the aircraft gains a significant amount of altitude, prompting the simulator to issue a strong correction using the elevator. 
This abrupt correction induces a negative G-force condition, which is more physiologically demanding for pilots than positive G-force. 
In comparison with $s_6$ and $\varphi^c$, $s_4$ successfully reaches the target altitude without safety violations, and $s_5$ ascends more gradually, maintaining compliance with all safety thresholds and achieving the target with a probability of 1. 
This experiment highlights how environmental noise, such as turbulence, can drive the system into failure-inducing states, demonstrating the importance of accounting for such uncertainties in the analysis of safety failures within CPS.

\section{Related Work}
\label{sec:related}

Causal analysis is of growing importance in formal verification in hopes of answering the question of ``why?'' when it comes to faults in complex systems.
A survey written by Baier et al.\cite{baier2021verificationcausalitybasedexplications} reviews numerous approaches that are inspired by the Halpern-Pearl (HP) causality framework to explain system behaviors.
Recent research applies temporal logic to model and explain causality and bugs \cite{cdffhhms22,fk17,cffhms22,bffs23}. 
In the CPS domain and robotics, causality has been explored to repair AI‑enabled controllers via HP models, search algorithms, and constraint verification \cite{lrcsl23,maldonado2025robotpouringidentifyingcauses, 10.1007/978-3-031-24670-8_9, 10.3389/frobt.2024.1123762}. 
%
% Work has also been done on applying causality to computer vision, in particular image classification in \cite{KR2024-25,Chockler_2021_ICCV}, and attempt to explain image classifications.
%
The studies \cite{gs20,ga14,gss17,wg15,gmflx13,b24,gmr10} present alternative formal frameworks to the HP causal model, whereas our approach uses causal analysis specifically to pinpoint the cause of a given effect.
Yet, these works often address only the modeling aspects or overlook the scalability challenges in automated, counterfactual reasoning. 
A recent approach presented in \cite{10745858} offers a more efficient method for directly identifying failure inducing causes from system execution traces.
Furthermore, causal analysis is also applied in fields like medicine, where it helps uncover the causes of biological phenomena~\cite{GUHA2025161134,Rafieioskouei2025}.
Although methods have been proposed to explain counterexamples in model checking \cite{cdffhhms22,ishar2009}, our work specifically targets efficient failure cause identification in embedded systems.
Expanding from deterministic to probabilistic systems, such as probabilistic programs and Markov chains, demands adaptations of causality notions to account for event likelihoods. 
In \cite{Ziemek2022}, Ziemek et al. use a different notion of causality extended to Markov Chains where the cause, which is composed of a set of executions, must cover all instances of the effect.
In \cite{10.1007/978-3-030-99253-8_3}, Baier et al. introduce the notion of probability-raising causation in MDPs, and extended this idea to explore quality measures of predictors and the notion of probability-raising policies as schedulers in MDPs \cite{baier2024formalqualitymeasurespredictors}.
\camera{While their framework allows for cycles, our work explores a different dimension of causality that preserves contingencies between the actual and counterfactual worlds, rather than focusing solely on probability-raising causes.}
In parallel to the above formal verification-oriented approach, Kleinberg et al. introduce actual causality in Markov chains from a data-driven perspective \cite{10.5555/1795114.1795150}. This perspective was later extended to token causality in \cite{10.5555/3031748.3031826} and further extended in \cite{DBLP:conf/flairs/ZhengK17}.
In \cite{maldonado2025robotpouringidentifyingcauses} the authors explore the probabilistic setting, but do so from the perspective of robot task execution.
They derive probability distributions from a simulation in which a robot is tasked with pouring one container of marbles into another container.
Their goal is to identify causes of unwanted behavior and to evaluate corrective actions the robotic agent can take. However we aim to propose an efficient method to discover causes, specifically in DTMCs.
\section{Conclusion and Future Work}
\label{sec:concl}

\vmcai{In this paper, we proposed two algorithms for efficient discovery of probabilistic actual causes due to Fenton-Glynn in systems characterized by stochastic behavior and noise.}
We (1) formulated the discovery of probabilistic actual causality in computing systems as an SMT problem, and (2) addressed the scalability challenges by introducing an abstraction-refinement technique that significantly improves efficiency. 
We demonstrated the effectiveness of our approach through three case studies, identifying probabilistic causes of safety violations in (1) the Mountain Car problem, (2) the Lunar Lander benchmark, and (3) MPC controller for an F-16 autopilot simulator.

This paper is the first step in formal analysis of actual causality in probabilistic systems. There are two immediate extensions with significant practical implications: actual causal inference in input models that allow nondeterministic decision making (MDPs) and those where full observability is not possible (POMDPs). In this paper, we formalize PAC in the hyperproperty setting. In \cite{hsu2025hyprl}, reinforcement learning is used to generate policies that maximize the satisfaction of hyperproperties. Combining these two studies could enable training agents that learn to identify counterfactual scenarios or achieve counterfactual realizability \cite{raghavan2025counterfactual}.

\bibliography{bibliography}

@preamble{"\newcommand{\SortNoop}[1]{}"}

@string{toplas = "ACM Transactions on Programming Languages and Systems"}

@inproceedings{ab18,
  author    = {E. {\'{A}}brah{\'{a}}m and B. Bonakdarpour},
  title     = {{H}yper{PCTL}: {A} Temporal Logic for Probabilistic 
Hyperproperties},
  booktitle = {Proceedings of the 15th International Conference on 
Quantitative Evaluation of Systems (QEST)},
  pages     = {20--35},
  year      = {2018}
}

@InProceedings{ap,
title={A Logical View of Composition and Refinement},
author={Mart{\'\i}n Abadi and Gordon D. Plotkin},
crossref={popl18},
pages={323--332},
source={http://theory.lcs.mit.edu/~dmjones/hbp/bibs/ley/popl/popl.bib}
}

@Article{aw,
  author =   "T. Anderson and R. W. Witty",
  title =    "Safe Programming",
  OPTcrossref =  "",
  OPTkey =   "",
  journal =      "BIT",
  year =     "1978",
  volume =   "18",
  OPTnumber =    "",
  pages =    "1-8",
  OPTmonth =     "",
  OPTnote =      "",
  OPTannote =    ""
}

@Article{f,
  author =   {A. Fekete},
  title =    {Formal models of communication services: {A} case study},
  journal =      {IEEE Computer},
  year =     {1993},
  OPTkey =   {},
  OPTvolume =    {},
  OPTnumber =    {},
  month =    {August},
  pages =    {37-47},
  OPTnote =      {},
  OPTannote =    {}
}

@Article{h,
title={Wait-Free Synchronization},
author={Maurice Herlihy},
journal=toplas,
pages={124--149},
month=jan,
year=1991,
volume=13,
number=1,
source={http://theory.lcs.mit.edu/~dmjones/hbp/bibs/ley/toplas/toplas.bib}
}

@Article{l,
  author =   {C. A. Lee},
  title =    {Barrier synchronization. over multistage interconnection networks},
  journal =      {IEEE Symposium on Parallel and Distributed Processing},
  year =     {1990},
  OPTkey =   {},
  OPTvolume =    {},
  OPTnumber =    {},
  month =    {December},
  pages =    {130-133},
  OPTnote =      {},
  OPTannote =    {}
}

@string{toplas = {ACM Transactions on Programming Languages and
Systems}}

@Article{m,
  author =   "Hector Garcia-Molina",
  title =    "Failure in distributed computing election of a coordinator for configuration",
  OPTcrossref =  "",
  OPTkey =   "",
  journal =      "IEEE Transactions on Computers",
  year =     "1982",
  volume =   "31",
  OPTnumber =    "",
  pages =    "148-159",
  OPTmonth =     "",
  OPTnote =      "",
  OPTannote =    ""
}

@Article{p,
  author =   {D. K. Panda},
  title =    {Fast barrier Synchronization in Wormhole k-ary n-cube networks with multidestination works},
  journal =      {Special Issue on High Performance Computer Architecture of the Journal of Future Generation Computer Systems (FGCS)},
  year =     {1995},
  OPTkey =   {},
  OPTvolume =    {11},
  OPTnumber =    {},
  OPTmonth =     {},
  OPTpages =     {585--602},
  OPTnote =      {},
  OPTannote =    {}
}

@Proceedings{popl18,
title={Conference Record of the Eighteenth Annual ACM Symposium on
Principles of Programming Languages},
booktitle={Conference Record of the Eighteenth Annual ACM Symposium on
Principles of Programming Languages},
month=jan,
year=1991,
address={Orlando, Florida},
c-organization={ACM},
key={ACM},
crossrefonly=1,
source={http://theory.lcs.mit.edu/~dmjones/hbp/bibs/ley/popl/popl.bib}
}

@Article{r,
  author =   "K. Raymond",
  title =    "A tree based algorithm for mutual exclusion",
  OPTcrossref =  "",
  OPTkey =   "",
  journal =      "ACM Transactions on Computer Systems",
  year =     "1989",
  volume =   "7",
  OPTnumber =    "",
  pages =    "61-77",
  OPTmonth =     "",
  OPTnote =      "",
  OPTannote =    ""
}

@PhdThesis{s,
  author =   {H. Schepers},
  title =    {Fault Tolerance and Timing of Distributed Systems: Compositional specification and verification},
  school =   {Eindhoven University},
  year =     {1994},
  OPTkey =   {},
  OPTaddress =   {},
  OPTtype =      {},
  OPTmonth =     {},
  OPTnote =      {},
  OPTannote =    {}
}

@PhdThesis{v,
  author =   "G. Varghese",
  title =    "Self-stabilization by local checking and correction",
  school =   "MIT/LCS/TR-583",
  year =     "1993",
  OPTcrossref =  "",
  OPTkey =   "",
  OPTaddress =   "",
  OPTmonth =     "",
  OPTtype =      "",
  OPTnote =      "",
  OPTannote =    ""
}

@Article{w,
  author =   {D. Weber},
  title =    {Formal Specification of fault-tolerance and its relation to computer security},
  journal =      {ACM Software Engineering Notes},
  year =     {1989},
  OPTkey =   {},
  volume =   {14},
  number =   {3},
  OPTmonth =     {},
  pages =    {273-277},
  OPTnote =      {},
  OPTannote =    {}
}

@Book{bk08-book,
 author = {C. Baier and J-.P. Katoen},
 title = {Principles of Model Checking},
 year = {2008},
 publisher = {The MIT Press},
}

@inproceedings{dmb08,
  author    = {L. M. de Moura and N. Bj{\o}rner},
  title     = {Z3: An Efficient {SMT} Solver},
  booktitle = {Tools and Algorithms for the Construction and Analysis of Systems (TACAS)},
  year      = {2008},
  pages     = {337-340}
}

@article{pac,
  author    = {M. Patrignani and A. Ahmed and D. Clarke},
  title     = {Formal Approaches to Secure Compilation},
  journal   = {{ACM} Computing Surveys},
  volume    = {1},
  number    = {1},
  year      = {2018},
  month     = {January},
  pages     = {1.1 -- 1.46}
}

@book{h16,
  author    = {J. Y. Halpern},
  title     = {Actual Causality},
  publisher = {{MIT} Press},
  year      = {2016},
  isbn      = {978-0-262-03502-6}
}

@inproceedings{ga14,
  author       = {G. Goessler and L. Astefanoaei},
  title        = {Blaming in component-based real-time systems},
  booktitle    = {Proceedings of the International Conference on Embedded Software (EMSOFT)},
  pages        = {7:1--7:10},
  publisher    = {{ACM}},
  year         = {2014}
}

@article{gs20,
  author       = {G. G{\"{o}}ssler and J.{-}B. Stefani},
  title        = {Causality analysis and fault ascription in component-based systems},
  journal      = {Theoretical Computer Science},
  volume       = {837},
  pages        = {158--180},
  year         = {2020}
}

@inproceedings{gss17,
  author       = {G. G{\"{o}}{\ss}ler and O. Sokolsky and J.{-}B. Stefani},
  title        = {Counterfactual Causality from First Principles?},
  booktitle    = {Proceedings 2nd International Workshop on Causal Reasoning for Embedded and safety-critical Systems Technologies (CREST)},
  volume       = {259},
  pages        = {47--53},
  year         = {2017},
}

@inproceedings{wg15,
  author       = {S. Wang and Y. Geoffroy and 
  					G. G{\"{o}}{\ss}ler and O. Sokolsky and I. Lee},
  title        = {A Hybrid Approach to Causality Analysis},
  booktitle    = {Proceedings of the 6th International Conference Runtime Verification (RV)},
  pages        = {250--265},
  publisher    = {Springer},
  year         = {2015}
}

@inproceedings{gmflx13,
  author       = {G. G{\"{o}}{\ss}ler and
                  D. Le M{\'{e}}tayer},
  editor       = {J. L. Fiadeiro and
                  Z. Liu and
                  J. Xue},
  title        = {A General Trace-Based Framework of Logical Causality},
  booktitle    = {Proceedings of the10th International Symposium on Formal Aspects of Component Software (FACS)},
  pages        = {157--173},
  year         = {2013}
}

@inproceedings{gmr10,
  author       = {G. G{\"{o}}{\ss}ler and
                  D. Le M{\'{e}}tayer and
                  J.{-}B. Raclet},
  title        = {Causality Analysis in Contract Violation},
  booktitle    = {Proceedings of the First International Conference on Runtime Verification (RV)},
  pages        = {270--284},
  year         = {2010}
}

@inproceedings{bffs23,
  author       = {R. Beutner and B. Finkbeiner and
                  H. Frenkel and J. Siber},
  title        = {Checking and Sketching Causes on Temporal Sequences},
  booktitle    = {Proceedings of the 21st International
                  Symposium Automated Technology for Verification and Analysis (ATVA)},
  pages        = {314--327},
  year         = {2023},
}

@inproceedings{cffhms22,
  author       = {N. Coenen and B. Finkbeiner and H. Frenkel and
                  C. Hahn and N. Metzger and J. Siber},
  title        = {Temporal Causality in Reactive Systems},
  booktitle    = {Proceedings of the 20th International
                  Symposium on Automated Technology for Verification and Analysis (ATVA)},
  pages        = {208--224},
  publisher    = {Springer},
  year         = {2022},
}

@inproceedings{fk17,
  author       = {B. Finkbeiner and A. Kupriyanov},
  title        = {Causality-based Model Checking},
  booktitle    = {Proceedings 2nd International Workshop on Causal Reasoning for Embedded and safety-critical Systems Technologies (CREST)},
  volume       = {259},
  pages        = {31--38},
  year         = {2017},
}

@inproceedings{cdffhhms22,
  author       = {N. Coenen and R. Dachselt and B. Finkbeiner and H. Frenkel and
                  C. Hahn and T. Horak and N. Metzger and J. Siber},
  title        = {Explaining Hyperproperty Violations},
  booktitle    = {Proceedings of the 34th International Conference on Computer Aided Verification(CAV), Part {I}},
  pages        = {407--429},
  year         = {2022},
}

@article{fg17,
	title = {A {Proposed} {Probabilistic} {Extension} of the {Halpern} and {Pearl} {Definition} of ‘{Actual} 
	{Cause}’},
	volume = {68},
	number = {4},
	journal = {The British Journal for the Philosophy of Science},
	author = {L. Fenton-Glynn},
	month = dec,
	year = {2017},
	pmid = {29593362},
	pmcid = {PMC5865829},
	pages = {1061--1124},
}

@misc{openaigym,
  Author = {G. Brockman and V. Cheung and L. Pettersson and J. Schneider and J. Schulman and J. 
  Tang and W. Zaremba},
  Title = {{O}pen{AI} {G}ym},
  Year = {2016},
  Eprint = {arXiv:1606.01540},
}

@inproceedings{ARCH18:Verification_Challenges_in_F_16,
  author    = {P. Heidlauf and A. Collins and M. Bolender and S. Bak},
  title     = {Verification Challenges in {F}-16 Ground Collision Avoidance and Other Automated 
  Maneuvers},
  booktitle = {ARCH18. 5th International Workshop on Applied Verification of Continuous and Hybrid Systems},
  volume    = {54},
  pages     = {208--217},
  year      = {2018}
}

@article{stable-baselines3,
  author  = {Antonin Raffin and Ashley Hill and Adam Gleave and Anssi Kanervisto and Maximilian Ernestus and Noah Dormann},
  title   = {Stable-Baselines3: Reliable Reinforcement Learning Implementations},
  journal = {Journal of Machine Learning Research},
  year    = {2021},
  volume  = {22},
  number  = {268},
  pages   = {1-8},
  url     = {http://jmlr.org/papers/v22/20-1364.html}
}

@misc{rl-zoo3,
  author = {Raffin, Antonin},
  title = {RL Baselines3 Zoo},
  year = {2020},
  publisher = {GitHub},
  journal = {GitHub repository},
  howpublished = {\url{https://github.com/DLR-RM/rl-baselines3-zoo}},
}

@ARTICLE{10745858,
  author={Rafieioskouei, Arshia and Bonakdarpour, Borzoo},
  journal={IEEE Transactions on Computer-Aided Design of Integrated Circuits and Systems}, 
  title={Efficient Discovery of Actual Causality Using Abstraction Refinement}, 
  year={2024},
  volume={43},
  number={11},
  pages={4274-4285},
}

@misc{baier2021verificationcausalitybasedexplications,
      title={From Verification to Causality-based Explications}, 
      author={Christel Baier and Clemens Dubslaff and Florian Funke and Simon Jantsch and Rupak Majumdar and Jakob Piribauer and Robin Ziemek},
      year={2021},
      eprint={2105.09533},
      archivePrefix={arXiv},
      primaryClass={cs.LO},
}

@INPROCEEDINGS{7243738,

  author={Datta, Anupam and Garg, Deepak and Kaynar, Dilsun and Sharma, Divya and Sinha, Arunesh},

  booktitle={2015 IEEE 28th Computer Security Foundations Symposium}, 

  title={Program Actions as Actual Causes: A Building Block for Accountability}, 

  year={2015},

  volume={},

  number={},

  pages={261-275},
}

@misc{maldonado2025robotpouringidentifyingcauses,
      title={Robot Pouring: Identifying Causes of Spillage and Selecting Alternative Action Parameters Using Probabilistic Actual Causation}, 
      author={Jaime Maldonado and Jonas Krumme and Christoph Zetzsche and Vanessa Didelez and Kerstin Schill},
      year={2025},
      eprint={2502.09395},
      archivePrefix={arXiv},
      primaryClass={cs.RO},
      url={https://arxiv.org/abs/2502.09395}, 
}

@Article{Rafieioskouei2025,
author={Rafieioskouei, Arshia
and Rogale, Kenneth
and Saei, Amir Ata
and Mahmoudi, Morteza
and Bonakdarpour, Borzoo},
title={Beyond Correlation: Establishing Causality in Protein Corona Formation for Nanomedicine},
journal={Molecular Pharmaceutics},
year={2025},
month={Apr},
day={09},
publisher={American Chemical Society},
issn={1543-8384},
}

@article{GUHA2025161134,
title = {AI-driven prediction of cardio-oncology biomarkers through protein corona analysis},
journal = {Chemical Engineering Journal},
volume = {509},
pages = {161134},
year = {2025},
author = {Avirup Guha and Seyed Amirhossein Sadeghi and Harikrishnan Hyma Kunhiraman and Fei Fang and Qianyi Wang and Arshia Rafieioskouei and Shaun Grumelot and Hassan Gharibi and Amir Ata Saei and Maryam Sayadi and Neal L. Weintraub and Sachi Horibata and Phillip Chung-Ming Yang and Borzoo Bonakdarpour and Mohammad Ghassemi and Liangliang Sun and Morteza Mahmoudi},
issn = {1385-8947},
}

@InProceedings{10.1007/978-3-030-99253-8_3,
author="Baier, Christel
and Funke, Florian
and Piribauer, Jakob
and Ziemek, Robin",
editor="Bouyer, Patricia
and Schr{\"o}der, Lutz",
title="On probability-raising causality in Markov decision processes",
booktitle="Foundations of Software Science and Computation Structures",
year="2022",
publisher="Springer International Publishing",
address="Cham",
pages="40--60",
}

@article{b24,
	author       = {M. Broy},
	title        = {Time, causality, and realizability: Engineering interactive, distributed
	software systems},
	journal      = {Journal of Systems and Software},
	volume       = {210},
	pages        = {111940},
	year         = {2024}
}

@inproceedings{lrcsl23,
	author       = {P. Lu and I. Ruchkin and	  M. Cleaveland and
	Oleg Sokolsky and I. Lee},
	title        = {Causal Repair of Learning-Enabled Cyber-Physical Systems},
	booktitle    = {Proceedings of the {IEEE} International Conference on Assured Autonomy (ICAA)},
	pages        = {1--10},
	year         = {2023}
}

@inproceedings{ishar2009,
  author       = {I. Beer and
                  S. Ben{-}David and
                  H. Chockler and
                  A. Orni and
                  R. J. Trefler},
  title        = {Explaining Counterexamples Using Causality},
  booktitle    = {Proceedings of the 21st International Conference on Computer Aided Verification 
  (CAV)},
  volume       = {5643},
  pages        = {94--108},
  year         = {2009}
}

@InProceedings{finkbeiner_et_al:LIPIcs.FSTTCS.2024.22,
  author =	{Finkbeiner, Bernd and Jahn, Felix and Siber, Julian},
  title =	{{Counterfactual Explanations for MITL Violations}},
  booktitle =	{44th IARCS Annual Conference on Foundations of Software Technology and Theoretical Computer Science (FSTTCS 2024)},
  pages =	{22:1--22:25},
  series =	{Leibniz International Proceedings in Informatics (LIPIcs)},
  ISBN =	{978-3-95977-355-3},
  ISSN =	{1868-8969},
  year =	{2024},
  volume =	{323},
  editor =	{Barman, Siddharth and Lasota, S{\l}awomir},
  publisher =	{Schloss Dagstuhl -- Leibniz-Zentrum f{\"u}r Informatik},
  address =	{Dagstuhl, Germany},
  URN =		{urn:nbn:de:0030-drops-222116},
}

@inproceedings{10.5555/1795114.1795150,
author = {Kleinberg, Samantha and Mishra, Bud},
title = {The temporal logic of causal structures},
year = {2009},
isbn = {9780974903958},
publisher = {AUAI Press},
address = {Arlington, Virginia, USA},
booktitle = {Proceedings of the Twenty-Fifth Conference on Uncertainty in Artificial Intelligence},
pages = {303–312},
numpages = {10},
location = {Montreal, Quebec, Canada},
series = {UAI '09}
}

@inproceedings{10.5555/3031748.3031826,
author = {Kleinberg, Samantha and Mishra, Bud},
title = {The temporal logic of token causes},
year = {2010},
isbn = {1577354516},
publisher = {AAAI Press},
booktitle = {Proceedings of the Twelfth International Conference on Principles of Knowledge Representation and Reasoning},
pages = {575–577},
numpages = {3},
location = {Toronto, Ontario, Canada},
series = {KR'10}
}

@inproceedings{DBLP:conf/flairs/ZhengK17,
  author       = {Min Zheng and
                  Samantha Kleinberg},
  editor       = {Vasile Rus and
                  Zdravko Markov},
  title        = {A Method for Automating Token Causal Explanation and Discovery},
  booktitle    = {Proceedings of the Thirtieth International Florida Artificial Intelligence
                  Research Society Conference, {FLAIRS} 2017, Marco Island, Florida,
                  USA, May 22-24, 2017},
  pages        = {176--181},
  publisher    = {{AAAI} Press},
  year         = {2017},
}

@article{Ziemek2022,
author={Ziemek, Robin
and Piribauer, Jakob
and Funke, Florian
and Jantsch, Simon
and Baier, Christel},
title={Probabilistic causes in Markov chains},
journal={Innovations in Systems and Software Engineering},
year={2022},
volume={18},
number={3},
pages={347-367},
}

@misc{baier2024formalqualitymeasurespredictors,
      title={Formal Quality Measures for Predictors in Markov Decision Processes}, 
      author={Christel Baier and Sascha Klüppelholz and Jakob Piribauer and Robin Ziemek},
      year={2024},
      eprint={2412.11754},
      archivePrefix={arXiv},
      primaryClass={cs.LO}, 
}

@InProceedings{10.1007/978-3-540-70545-1_16,
author="Hermanns, Holger
and Wachter, Bj{\"o}rn
and Zhang, Lijun",
editor="Gupta, Aarti
and Malik, Sharad",
title="Probabilistic {CEGAR}",
booktitle="Computer Aided Verification",
year="2008",
publisher="Springer Berlin Heidelberg",
address="Berlin, Heidelberg",
pages="162--175",
isbn="978-3-540-70545-1"
}

@InProceedings{10.1007/978-3-031-24670-8_9,
author="Araujo, Hugo
and Holthaus, Patrick
and Gou, Marina Sarda
and Lakatos, Gabriella
and Galizia, Giulia
and Wood, Luke
and Robins, Ben
and Mousavi, Mohammad Reza
and Amirabdollahian, Farshid",
editor="Cavallo, Filippo
and Cabibihan, John-John
and Fiorini, Laura
and Sorrentino, Alessandra
and He, Hongsheng
and Liu, Xiaorui
and Matsumoto, Yoshio
and Ge, Shuzhi Sam",
title="Kaspar Causally Explains",
booktitle="Social Robotics",
year="2022",
publisher="Springer Nature Switzerland",
address="Cham",
pages="85--99",
isbn="978-3-031-24670-8"
}

@article{10.3389/frobt.2024.1123762,
author={Zibaei, Ehsan  and Borth, Robin },
title={Building causal models for finding actual causes of unmanned aerial vehicle failures},        
journal={Frontiers in Robotics and AI},        
volume={Volume 11 - 2024},
year="2024",
ISSN={2296-9144},
}

@inproceedings{
hsu2025hyprl,
title={{HYPRL}: Reinforcement Learning of Control Policies for Hyperproperties},
author={Tzu-Han Hsu and Arshia Rafieioskouei and Borzoo Bonakdarpour},
booktitle={The Thirty-ninth Annual Conference on Neural Information Processing Systems},
year={2025}
}

@inproceedings{
raghavan2025counterfactual,
title={Counterfactual Realizability},
author={Arvind Raghavan and Elias Bareinboim},
booktitle={The Thirteenth International Conference on Learning Representations},
year={2025},
url={https://openreview.net/forum?id=uuriavczkL}
}
\bibliographystyle{abbrv}

\newpage
\appendix
\section{Proof}
\begin{proof}
\label{app:proof}
Let  $\hat{\dtmc}$ is the abstract model of $\dtmc$.
Formally, we need to prove the following.
{\bf Assumption:} Let us assume that:
\begin{align*}
\nonumber   \exists \hat{\sigma}. \forall \hat{\sigma}'. & \mathbb{P}^{\text{min}} \Big(\neg \varphi^e_{\hat{\sigma}}~\U~ (\varphi_{\hat{\sigma}}^c \land \mathbb{P}^{\text{min}}_{>0}(\F 
	\varphi_{\hat{\sigma}}^e))\Big) >  \mathbb{P}^{\text{max}}\Big(\neg \varphi_{\hat{\sigma}'} ^c \U 
	 \varphi_{\hat{\sigma}'}^e\Big) ~\land   \\ & \mathbb{P}_{=1}^{\min} \big(\bigwedge_{p\in W} \G (p_{\hat{\sigma}} \leftrightarrow 
		p_{\hat{\sigma}'})\big) && 
\end{align*}
{\bf Goal:} We should prove that:

\begin{align*}	\nonumber \varphi_{\pac} \triangleq \exists \sigma. \forall \sigma'. & \mathbb{P}\Big(\neg 
	\varphi^e_{\sigma}~\U~ (\varphi_{\sigma}^c \land \mathbb{P}_{>0}(\F 
	\varphi_{\sigma}^e))\Big)> 
	 \mathbb{P}\Big(\neg \varphi_{\sigma'} ^c \U 
	 \varphi_{\sigma'}^e\Big) \wedge \\
		& \label{eq:pac} \hspace{-3mm}\mathbb{P}_{=1} \big(\bigwedge_{p\in W} \G (p_{\sigma} \leftrightarrow 
		p_{\sigma'})\big)
\end{align*}
First, we acknowledge from \Cref{lemma:1} that the abstract model simulates the concrete model, and any reachability property satisfied by $\hat{\dtmc}$ is also satisfied by $\dtmc$. Next, we proceed as follows.
	\begin{itemize}
        \item We begin by addressing the first part of the $\varphi_\pac$ formula, which corresponds to satisfying the \textbf{PC1} condition. Assume that in the abstract model $\hat{\dtmc}$, we identify an abstract initial state $\hat{\sigma}$ in the actual world such that the minimum probability of reaching the effect holds: $\mathbb{P}^{\text{min}}_{>0}(\F \varphi_{\hat{\sigma}}^e)$. This implies that, even under the most pessimistic scheduler, the probability of eventually reaching a state satisfying $\varphi^e$ is strictly greater than zero. Since $\hat{\dtmc}$ is derived through abstraction from the concrete model $\dtmc$, each action in the abstract state corresponds to concrete state. Therefore, the existence of such a scheduler in $\hat{\dtmc}$ guarantees the existence of a corresponding path in $\dtmc$ that achieves a non-zero probability of reaching $\varphi^e$ in the actual world. Thus, the PC1 condition is satisfied in the concrete model as well.
        \item Next, we address the second part of the $\varphi_\pac$ formula, which corresponds to satisfying the \textbf{PC2} condition. Assume that in the abstract model $\hat{\dtmc}$, the minimum probability of reaching the effect $\varphi^e$ through the candidate cause $\varphi^c$ in the actual world $\hat{\sigma}$ is greater than the maximum probability of reaching $\varphi^e$ in the counterfactual world $\hat{\sigma}'$, where $\varphi^c$ is not realized. Formally, we assume:
$$\mathbb{P}^{\text{min}} \Big(\neg \varphi^e_{\hat{\sigma}}~\U~ (\varphi_{\hat{\sigma}}^c \land \mathbb{P}^{\text{min}}_{>0}(\F 
	\varphi_{\hat{\sigma}}^e))\Big) >  \mathbb{P}^{\text{max}}\Big(\neg \varphi_{\hat{\sigma}'} ^c \U 
	 \varphi_{\hat{\sigma}'}^e\Big)$$
This means that, even under the most pessimistic scheduler in the actual world $\hat{\sigma}$, the probability of eventually reaching a state satisfying $\varphi^e$ via $\varphi^c$ is strictly greater than the probability of reaching $\varphi^e$ under the most optimistic scheduler in the counterfactual world $\hat{\sigma}'$. As in the first part of the proof, since actions in the abstract model $\hat{\dtmc}$ correspond to concrete transitions in the underlying model $\dtmc$, the computation tree induced by the pessimistic scheduler in $\hat{\sigma}$ is realizable in the concrete model. Similarly, the computation tree induced by the optimistic scheduler in $\hat{\sigma}'$ is also concretizable. Therefore, there exists a computation tree in the actual world in concrete model $\dtmc$ whose probability of reaching the effect is strictly greater than that of any counterfactual computation trees in which $\varphi^c$ does not occur. This satisfies second part of our proof.

        \item Next, we prove that if the actual world computation tree $\hat{\sigma}$ and the counterfactual world computation tree $\hat{\sigma}'$ agree on all propositions in $W$ in the abstract model $\hat{\dtmc}$, i.e., 
\[
\mathbb{P}_{=1}^{\min} \left(\bigwedge_{p \in W} \G (p_{\hat{\sigma}} \leftrightarrow p_{\hat{\sigma}'})\right),
\]
then the corresponding actual and counterfactual computation trees $\sigma$ and $\sigma'$ in the concrete model $\dtmc$ also agree on all propositions in $W$. In \Cref{sec:pred}, we distinguish between two strategies for handling $W$. Below, we prove this property for both approaches:

\begin{itemize}
    \item \textbf{Enumerating Subgraphs.} If we enumerate subgraphs such that all paths within each subgraph are mutually stutter-equivalent with respect to $W$, then it is immediate that $\hat{\sigma}$ and $\hat{\sigma}'$ are stutter-equivalent with respect to $W$ by construction. Therefore, their corresponding computation trees in the concrete model also agree on $W$.
    
    \item \textbf{$W$-Preserving Abstraction.} Suppose we use a $W$-preserving abstraction function, and we find actual and counterfactual computation trees $\hat{\sigma}$ and $\hat{\sigma}'$ in the abstract model that are stutter-equivalent with respect to $W$. Since the encoding of $\SEABS$ checks all paths induced by all schedulers in the actual and counterfactual worlds, if in the abstraction we find an actual and a counterfactual world where the PAC condition holds, we can conclude that all possible paths in both worlds are stutter-trace equivalent with respect to the variables in $W$.
\end{itemize}

       \end{itemize} 
This concludes the proof.

\end{proof}

\section{Details on Enumerating Subgraphs and Contingency Variables}
\label{app:subgraph}
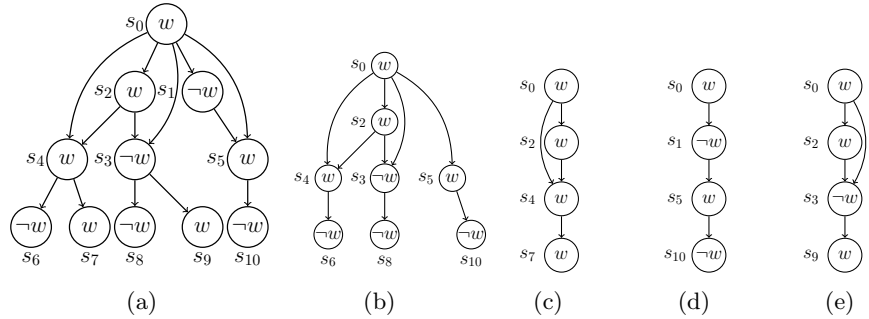
\begin{figure*}[t]
    \centering
    \begin{subfigure}[t]{0.3\textwidth}
        \centering
       \scalebox{0.6}{\begin{tikzpicture}[->, auto, thick,node distance=1.5cm]
                \Large
			\tikzstyle{every state}=[circle,draw=black, text=black,minimum size=0.9cm,inner sep=0.01cm]
			
			\node[state]    (s0) [circle]             {$w$};
			
			\node[state]    (s1)          [xshift= 0.8cm,below of=s0]      {$\neg w$};
			\node[state]    (s2)            [left of=s1]    {$w$};

			\node[state]   (s3)          [below of=s2]      {$\neg w$};
			\node[state]   (s4)            [left of=s3]    {$w$};
			
			\node[state]   (s5)          [xshift= 1cm, below of=s1]      {$w$};

			\node[state]    (s6)          [xshift= -0.8cm, below of=s4]      {$\neg w$};
			\node[state]    (s7)            [xshift=-0.2cm,right of=s6]    {$w$};
			\node[state]    (s8)          [ below of=s3]      {$\neg w$};
			\node[state]    (s9)            [right of=s8]    {$w$};
			
			\node[state]    (s10)          [below of=s5]      {$\neg w$};
			
			\path
			
			(s0)
			edge [] node {}(s1)
			edge [] node{} (s2)
                edge [bend left] node{} (s5)
                edge [bend left] node{} (s3)
                edge [bend right] node{} (s4)

			(s2) edge []  node {}(s3)
			edge []  node{}(s4)
			
			(s1)
			edge [] node  {}(s5)

			(s4)
			edge []  node {} (s6)
			edge [] node  {}(s7)
			(s3)
			edge []  node{}(s8)
			edge [] node {}(s9)
			
			(s5)
			edge [] node {} (s10);

                \node[left of=s0, xshift=0.8cm] {$s_0$};
                \node[left of=s1, xshift=0.7cm] {$s_1$};
                \node[left of=s2, xshift=0.8cm] {$s_2$};
                \node[left of=s3, xshift=0.8cm] {$s_3$};
                \node[left of=s4, xshift=0.8cm] {$s_4$};
                \node[left of=s5, xshift=0.8cm] {$s_5$};
                
                \node[below of=s6, yshift=0.8cm] {$s_{6}$};
                \node[below of=s7, yshift=0.8cm] {$s_{7}$};
                \node[below of=s8, yshift=0.8cm] {$s_{8}$};
                \node[below of=s9, yshift=0.8cm] {$s_{9}$};
                \node[below of=s10, yshift=0.8cm] {$s_{10}$};

		\end{tikzpicture}}
        \caption{}
        \label{fig:dtmcst}
    \end{subfigure}
    \begin{subfigure}[t]{0.2\textwidth}
        \centering
        \scalebox{0.5}{\begin{tikzpicture}[->, auto, thick,node distance=1.5cm]
\Large
			\tikzstyle{every state}=[circle,draw=black, text=black,minimum size=0.7cm,inner sep=0.01cm]
			
			\node[state]    (s0) [circle]             {$w$};

			\node[state]    (s2)            [below of=s0]    {$w$};

			\node[state]   (s3)          [below of=s2]      {$\neg w$};
			\node[state]   (s4)            [left of=s3]    {$w$};
			
			\node[state]   (s5)          [xshift= 1cm, below of=s1]      {$w$};

			\node[state]    (s6)          [xshift=0cm, below of=s4]      {$\neg w$};
			\node[state]    (s8)          [ below of=s3]      {$\neg w$};
			
			\node[state]    (s10)          [xshift=0.5cm, below of=s5]      {$\neg w$};
			
			\path
			
			(s0)

			edge [] node{} (s2)
                edge [bend left] node{} (s5)
                edge [bend left] node{} (s3)
                edge [bend right] node{} (s4)

			(s2) edge []  node {}(s3)
			edge []  node{}(s4)

			(s4)
			edge []  node {} (s6)
			(s3)
			edge []  node{}(s8)
			
			(s5)
			edge [] node {} (s10);

                \node[left of=s0, xshift=0.8cm] {$s_0$};
                \node[left of=s2, xshift=0.8cm] {$s_2$};
                \node[left of=s3, xshift=0.8cm] {$s_3$};
                \node[left of=s4, xshift=0.8cm] {$s_4$};
                \node[left of=s5, xshift=0.8cm] {$s_5$};
                
                \node[below of=s6, yshift=0.8cm] {$s_{6}$};
                \node[below of=s8, yshift=0.8cm] {$s_{8}$};
                \node[below of=s10, yshift=0.8cm] {$s_{10}$};

		\end{tikzpicture}}
        \caption{}
        \label{fig:dtmcst1}
    \end{subfigure}
    \begin{subfigure}[t]{0.15\textwidth}
        \centering
        \scalebox{0.5}{\begin{tikzpicture}[->, auto, thick,node distance=1.5cm]
\Large
			\tikzstyle{every state}=[circle,draw=black, text=black,minimum size=0.9cm,inner sep=0.01cm]
			
			\node[state]    (s0) [circle]             {$w$};

			\node[state]    (s2)            [below of=s0]    {$w$};

			\node[state]   (s4)            [below of=s2]    {$w$};

			\node[state]    (s7)            [below of=s4]    {$w$};
			
			\path
			
			(s0)
			edge [] node{} (s2)
                edge [bend right] node{} (s4)

			(s2) 
			edge []  node{}(s4)

			(s4)
			edge [] node  {}(s7);

                \node[left of=s0, xshift=0.6cm] {$s_0$};
                \node[left of=s2, xshift=0.6cm] {$s_2$};
                \node[left of=s4, xshift=0.6cm] {$s_4$};

                \node[left of=s7, xshift=0.6cm] {$s_{7}$};

		\end{tikzpicture}}
        \caption{}
        \label{fig:dtmcst2}
    \end{subfigure}
    \begin{subfigure}[t]{0.15\textwidth}
        \centering
        \scalebox{0.5}{\begin{tikzpicture}[->, auto, thick,node distance=1.5cm]
			\tikzstyle{every state}=[circle,draw=black, text=black,minimum size=0.9cm,inner sep=0.01cm]
            \Large
			
			\node[state]    (s0) [circle]             {$w$};
			
			\node[state]    (s1)          [below of=s0]      {$\neg w$};

			\node[state]   (s5)          [below of=s1]      {$w$};
			
			\node[state]    (s10)          [below of=s5]      {$\neg w$};
			
			\path
			
			(s0)
			edge [] node {}(s1)

			(s1)
			edge [] node  {}(s5)

			(s5)
			edge [] node {} (s10);

                \node[left of=s0, xshift=0.6cm] {$s_0$};
                \node[left of=s1, xshift=0.6cm] {$s_1$};
                \node[left of=s5, xshift=0.6cm] {$s_5$};
              
                \node[left of=s10, xshift=0.6cm]  {$s_{10}$};

		\end{tikzpicture}}
        \caption{}
        \label{fig:dtmcst3}
    \end{subfigure}
    \begin{subfigure}[t]{0.15\textwidth}
        \centering
        \scalebox{0.5}{\begin{tikzpicture}[->, auto, thick,node distance=1.5cm]
\Large
			\tikzstyle{every state}=[circle,draw=black, text=black,minimum size=0.9cm,inner sep=0.01cm]
			
			\node[state]    (s0) [circle]             {$w$};

			\node[state]    (s2)            [below of=s0]    {$w$};

			\node[state]   (s3)          [below of=s2]      {$\neg w$};
		
			\node[state]    (s9)            [below of=s3]    {$w$};

			\path
			
			(s0)

			edge [] node{} (s2)

                edge [bend left] node{} (s3)

			(s2) edge []  node {}(s3)

			(s3)

			edge [] node {}(s9);

                \node[left of=s0, xshift=0.6cm] {$s_0$};

                \node[left of=s2, xshift=0.6cm] {$s_2$};
                \node[left of=s3, xshift=0.6cm] {$s_3$};
                
                \node[left of=s9, xshift=0.6cm] {$s_{9}$};

		\end{tikzpicture}
		}
        \caption{}
        \label{fig:dtmcst4}
    \end{subfigure}

    \caption{Process of generating subgraphs from a DTMC based on stutter-trace equivalence. (a) shows the full DTMC. Subfigures (b)–(e) represent subgraphs where all paths share the same collapsed trace over a variable of interest: (b) corresponds to the trace pattern $w, \neg w$; (c) to $w$; (d) to $w, \neg w, w, \neg w$; and (e) to $w, \neg w, w$.}
    \label{fig:dtmc-stutter}
\end{figure*}

Before delving into the enumerating subgraph strategy, we clarify why we do not explicitly check the equivalence of contingencies.
Recall from \Cref{eq:pac} that the actual and counterfactual computation trees are required to agree on a set of propositions (or, more generally, variables) $W$.
Formally, this condition targets paths that exhibit identical patterns of values for variables in $W$. 
However, due to potential differences in sampling frequencies or event occurrences, a direct one-to-one state-wise comparison between paths is not feasible.
To address this challenge, we adopt the notion of stutter-trace equivalence with respect to the variables in $W$~\cite{bk08-book}.
We enforce this equivalence by decomposing the original DTMC $\dtmc$ into subgraphs $\{\dtmc^1, \dots, \dtmc^n\}$, where each $\dtmc^i$ represents a subgraph in which all paths are mutually stutter-equivalent with respect to $W$. 
Since we work with acyclic DTMCs derived from execution logs, this decomposition depends on the structure of the underlying DAG. 
If the DTMC contains diamond-shaped structures, the number of paths can grow exponentially resulting in exponential complexity for subgraph enumeration. 
However, if the DAG resembles a tree, the number of paths remains polynomial making the decomposition tractable.
\Cref{fig:dtmc-stutter} illustrates this decomposition process, showing how the DTMC in \Cref{fig:dtmcst} is partitioned into subgraphs depicted in \Cref{fig:dtmcst,fig:dtmcst1,fig:dtmcst2,fig:dtmcst3,fig:dtmcst4}.

\section{Causal Analysis of the Lunar Lander Case Study}
\label{app:extra-causal}

\begin{figure}[h]
  \centering
  \resizebox{0.5\textwidth}{!}{%
        \begin{tikzpicture}[->, auto, thick,node distance=3cm]

\large
			
			\tikzstyle{every state}=[rectangle,draw=black, text=black,minimum size=0.5cm,inner sep=0.1cm]

			\node[state]    (s0) [rectangle]             {$(x=0,y =1.92, act= 0)$};

            \node[state]    (s1) [yshift=1cm,xshift=-3cm,below of=s0]   [rectangle]             {$(0.02,1.46, 0)$};

            \node[state]    (s2) [xshift=3cm,right of=s1]   [rectangle]             {$(0.52,1.41, 3)$};

            \node[state]    (s22) [xshift=-3cm,yshift=1cm,below of=s2]   [rectangle]             {$(0.52,1.2, 2)$};

            \node[state]    (s23) [xshift=-3cm,yshift=1cm,below of=s22]   [rectangle]             {$(0.44,1.24, 2)$};

            \node[state]    (s24) [xshift=3cm,right of=s23]   [rectangle]             {\textcolor{Green}{$(-0.1,0.3, 0)$}};

            \node[state]    (s25) [yshift=-1cm,below of=s22]   [rectangle]             {$(0.7,0.8, 0)$};

            \node[state]    (s26) [left of=s25]   [rectangle]             {\textcolor{Green}{$(0.40,0.20, 0)$}};

             \node[state]    (s27) [right of=s25]   [rectangle]             {\textcolor{Green}{$(-0.1,0.45, 0)$}};

             \node[state]    (s28) [yshift=1cm,below of=s25]   [rectangle]             {\textcolor{Blue}{$(1.32,0.6, 2)$}};

             \node[state]    (s29) [right of=s28]   [rectangle]             {$(1.22,0.7, 2)$};

             \node[state]    (s30) [left of=s28]   [rectangle]             {\textcolor{Green}{$(0.1,0.2, 0)$}};

             \node[state]    (sef) [xshift=3cm,below of=s28]   [rectangle]             {\huge \textcolor{red}{$\varphi^e$}};

             \node[state]    (snoeff) [xshift=-3cm,left of=sef]   [rectangle]             {\huge \textcolor{Green}{$\neg \varphi^e$}};

			% \node[state]    (s1)          [yshift=1.9cm,xshift=1.5cm,below of=s0]      {$(0.4,0.03, 0)$};
			% \node[state]    (s2)            [left of=s1]    {$(0.3,0.01, 1)$};

			% \node[state]   (s4)          [yshift=2cm,below of=s2]      {$(0.35,0.02, 1)$};
			% \node[state]   (s5)            [xshift=0cm,left of=s4]    {$(0.38,0.02, 1)$};
			
			% \node[state]   (s6)          [yshift=2cm, below of=s1]      {$(0.45,0.03, 1)$};

			% \node[state]    (s8)          [xshift=-0.5cm, yshift=2cm,below of=s5]      {\textcolor{Green}{$(0.61,0.02, 0)$}};
			% \node[state]    (s9)            [xshift=-0.9cm,right of=s8]    {\textcolor{red}{$(0.5,0.02, 0)$}};
			% \node[state]    (s10)          [yshift=2cm,xshift=1cm,below of=s4]      {\textcolor{Green}{$(0.6,0.0, 0)$}};
			
			% \node[state]    (s12)          [yshift=2cm,xshift=0cm,below of=s6]      {\textcolor{red}{$(0.52,0.02, 1)$}};
			% \node[state]    (s13)            [xshift=0cm,xshift=-0.2cm,right of=s12]    {\textcolor{Green}{$(0.62,0.04, 1)$}};

			\path

            (s0)
			edge [] node[swap] {0.12} (s1)
			edge [] node[] {0.88} (s2)

            (s2)
			edge [dashed] node[swap] {0.36} (s22)

            (s22)
			edge [] node[swap] {0.72} (s23)
            edge [] node {0.28} (s24)

            (s23)
			edge [] node {0.8}(s25)
            edge [] node[swap] {0.14} (s26)
            edge [] node {0.05} (s27)

            (s25)
			edge [] node {0.26}(s28)
            edge [] node[] {0.16} (s29)
            edge [] node[swap] {0.58} (s30)

            (s28)
			edge [dashed] node {1}(sef)

             (s29)
			edge [dashed] node {0.76}(sef)
            edge [dashed] node[swap] {0.24}(snoeff)

			;

                \node[left of=s0] {$s_0$};
                \node[left of=s2, xshift=1.5cm] {$s_2$};
                \node[left of=s1, xshift=1.5cm] {$s_1$};

                \node[left of=s22, xshift=1.2cm] {$s_{22}$};
                \node[left of=s23, xshift=1.2cm] {$s_{23}$};
                \node[left of=s24, xshift=1.2cm] {$s_{24}$};
                \node[left of=s25, xshift=1.6cm] {$s_{25}$};
                \node[left of=s26, xshift=1.2cm] {$s_{26}$};
                \node[left of=s27, xshift=1.4cm] {$s_{27}$};
                \node[left of=s28, xshift=1.5cm] {\textcolor{Blue}{$\mathbf{s_{28}}$}};
                 \node[left of=s29, xshift=1.5cm] {$s_{29}$};
                  \node[left of=s30, xshift=1.5cm] {$s_{30}$};

		\end{tikzpicture}
			
	%
  }
  \caption{A Lunar Lander Scenario}
  \label{fig:lun-CA}
\end{figure}
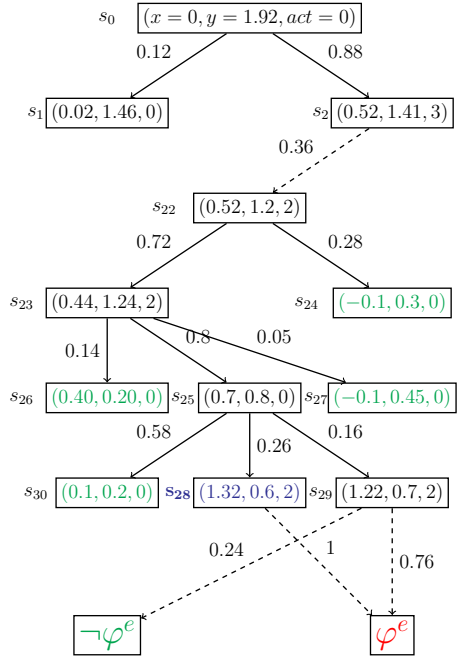

In \Cref{sec:lunar}, we discussed a scenario in which the Lunar Lander starts from an initial position $(x, y)$ and aims to land on the helipad located at $(0, 0) \pm \epsilon$, where we take $\epsilon = 0.5$. 
As shown in \Cref{fig:lun-CA}, the lander starts in state $s_0 \triangleq (x = 0 \land y = 1.92 \land \textit{vel}_x = 0.5 \land \textit{vel}_y = 3.6 \land \act = 0)$. Due to system noise, it eventually transitions into two possible successor states, $s_1$ and $s_2$. 
Our causal discovery algorithm identifies state $s_{28} \triangleq (x = 1.32 \land y = 0.6 \land \textit{vel}_x = -0.5 \land \textit{vel}_y = -0.7 \land \act = 2)$ as the cause of the failure. 
In contrast, in the counterfactual world, such as computation trees starting from $s_{29}$ and $s_{30}$, the probability of reaching the failure state is lower than in the actual world.
Specifically, the probabilities of reaching the effect from $s_{29}$ and $s_{30}$ are 0.76 and 0, respectively. 
Examining the system dynamics more closely, we observe that states $s_{28}$ and $s_{29}$ are quite similar. 
However, $s_{29}$ has a slightly higher altitude and an $x$-coordinate closer to the center of the helipad ($0 \pm \epsilon$), which enables the lander to recover from the rightward drift and land safely with probability 0.24. 
This example illustrates how even small perturbations or noise can lead to significantly different outcomes, including system failure.

\section{Hyperparameter Setting}
\label{app:hyp}

\begin{figure}[h!]
  \centering
  \begin{subfigure}[t]{0.48\textwidth}
    \centering
    \resizebox{\textwidth}{!}{\begin{tikzpicture}
  \begin{axis}[
      width=4\textwidth,
      height=4\textwidth,
      xlabel={\Huge \bfseries Split Ratio},
      ylabel={\Huge \bfseries  Time (s)},
      xtick={0,0.2,0.4,0.6,0.8,1},
      ytick={0.1},
      xlabel style={yshift=-1cm},
      ylabel style={yshift=1cm},
      xticklabel style={font=\Huge}, 
      yticklabel style={font=\Huge}, 
      ymode=log,
      ymajorgrids=true,
      yminorgrids=true,
      minor y tick num=4,              % 4 minor ticks per major step
      major grid style={gray!30},
      minor grid style={gray!10, dashed}, 
      legend style={
        at={(0.5,-0.15)},
        anchor=north,
        legend columns=5,
        font=\Huge,
        draw=black,
        inner sep=2pt
      },
    ]

    \addplot[thick, line width=2pt, yellow]
      table [x=SplitRatio, y=ARTime, col sep=comma] {images/lunrat7.csv};
    \addlegendentry{$|S|=78$}

    \addplot[thick, line width=2pt, orange]
      table [x=SplitRatio, y=ARTime, col sep=comma] {images/lunrat4.csv};
    \addlegendentry{$|S|=115$}

    \addplot[thick, line width=2pt, blue]
      table [x=SplitRatio, y=ARTime, col sep=comma] {images/lunrat1.csv};
    \addlegendentry{$|S|=192$}

    \addplot[thick, line width=2pt, red]
      table [x=SplitRatio, y=ARTime, col sep=comma] {images/lunrat2.csv};
    \addlegendentry{$|S|=212$}

    \addplot[thick, line width=2pt, Green]
      table [x=SplitRatio, y=ARTime, col sep=comma] {images/lunrat3.csv};
    \addlegendentry{$|S|=249$}

       \addplot[thick, line width=2pt, purple]
      table [x=SplitRatio, y=ARTime, col sep=comma] {images/lunrat6.csv};
    \addlegendentry{$|S|=291$}

        \addplot[thick, line width=2pt, cyan]
      table [x=SplitRatio, y=ARTime, col sep=comma] {images/lunrat8.csv};
    \addlegendentry{$|S|=468$}

        \addplot[thick, line width=2pt, olive]
      table [x=SplitRatio, y=ARTime, col sep=comma] {images/lunrat9.csv};
    \addlegendentry{$|S|=780$}

    \addplot[thick, line width=2pt, gray]
      table [x=SplitRatio, y=ARTime, col sep=comma] {images/lunrat5.csv};
    \addlegendentry{$|S|=879$}

     \addplot[thick, line width=2pt, magenta]
      table [x=SplitRatio, y=ARTime, col sep=comma] {images/lunrat10.csv};
    \addlegendentry{$|S|=1187$}

  \end{axis}
\end{tikzpicture}}
    \caption{Lunar Lander}
    \label{fig:lunrat}
  \end{subfigure}\hfill
  \begin{subfigure}[t]{0.48\textwidth}
    \centering
    \resizebox{\textwidth}{!}{\begin{tikzpicture}
  \begin{axis}[
      width=4\textwidth,
      height=4\textwidth,
      xlabel={\Huge \bfseries Split Ratio},
      ylabel={\Huge \bfseries  Time (s)},
      xtick={0,0.2,0.4,0.6,0.8,1},
      xlabel style={yshift=-1cm},
      ylabel style={yshift=1cm},
      xticklabel style={font=\Huge}, 
      yticklabel style={font=\Huge}, 
      ymode=log,
      ymajorgrids=true,
      yminorgrids=true,
      minor y tick num=4,              % 4 minor ticks per major step
      major grid style={gray!30},
      minor grid style={gray!10, dashed}, 
      legend style={
        at={(0.5,-0.15)},
        anchor=north,
        legend columns=5,
        font=\Huge,
        draw=black,
        inner sep=2pt
      },
    ]

    \addplot[thick, line width=2pt, yellow]
      table [x=SplitRatio, y=ARTime, col sep=comma] {images/carar7.csv};
    \addlegendentry{$|S|=78$}

    \addplot[thick, line width=2pt, orange]
      table [x=SplitRatio, y=ARTime, col sep=comma] {images/carar4.csv};
    \addlegendentry{$|S|=115$}

    \addplot[thick, line width=2pt, blue]
      table [x=SplitRatio, y=ARTime, col sep=comma] {images/carar1.csv};
    \addlegendentry{$|S|=192$}

    \addplot[thick, line width=2pt, red]
      table [x=SplitRatio, y=ARTime, col sep=comma] {images/carar2.csv};
    \addlegendentry{$|S|=212$}

    \addplot[thick, line width=2pt, Green]
      table [x=SplitRatio, y=ARTime, col sep=comma] {images/carar3.csv};
    \addlegendentry{$|S|=249$}

       \addplot[thick, line width=2pt, purple]
      table [x=SplitRatio, y=ARTime, col sep=comma] {images/carar6.csv};
    \addlegendentry{$|S|=291$}

        \addplot[thick, line width=2pt, cyan]
      table [x=SplitRatio, y=ARTime, col sep=comma] {images/carar8.csv};
    \addlegendentry{$|S|=468$}

        \addplot[thick, line width=2pt, olive]
      table [x=SplitRatio, y=ARTime, col sep=comma] {images/carar9.csv};
    \addlegendentry{$|S|=780$}

    \addplot[thick, line width=2pt, gray]
      table [x=SplitRatio, y=ARTime, col sep=comma] {images/carar5.csv};
    \addlegendentry{$|S|=879$}

     \addplot[thick, line width=2pt, magenta]
      table [x=SplitRatio, y=ARTime, col sep=comma] {images/carar10.csv};
    \addlegendentry{$|S|=1187$}

  \end{axis}
\end{tikzpicture}}
    \caption{Mountain Car}
    \label{fig:mcarrat}
  \end{subfigure}\hfill
  \begin{subfigure}[t]{0.48\textwidth}
    \centering
    \resizebox{\textwidth}{!}{\begin{tikzpicture}
  \begin{axis}[
      width=4\textwidth,
      height=4\textwidth,
      xlabel={\Huge \bfseries Split Ratio},
      ylabel={\Huge \bfseries Time (s)},
      xtick={0,0.2,0.4,0.6,0.8,1},
      xlabel style={yshift=-1cm},
      ylabel style={yshift=1cm},
      xticklabel style={font=\Huge}, 
      yticklabel style={font=\Huge}, 
      ymode=log,
      ymajorgrids=true,
      yminorgrids=true,
      minor y tick num=4,              % 4 minor ticks per major step
      major grid style={gray!30},
      minor grid style={gray!10, dashed}, 
      legend style={
        at={(0.5,-0.15)},
        anchor=north,
        legend columns=5,
        font=\Huge,
        draw=black,
        inner sep=2pt
      },
    ]
     \addplot[thick, line width=2pt, orange]
      table [x=SplitRatio, y=ARTime, col sep=comma] {images/f16rat4.csv};
    \addlegendentry{$|S|=84$}

    \addplot[thick, line width=2pt, blue]
      table [x=SplitRatio, y=ARTime, col sep=comma] {images/f16rat1.csv};
    \addlegendentry{$|S|=130$}

    \addplot[thick, line width=2pt, cyan]
      table [x=SplitRatio, y=ARTime, col sep=comma] {images/f16rat8.csv};
    \addlegendentry{$|S|=371$}

           \addplot[thick, line width=2pt, red]
      table [x=SplitRatio, y=ARTime, col sep=comma] {images/f16rat2.csv};
    \addlegendentry{$|S|=401$}

     \addplot[thick, line width=2pt, magenta]
      table [x=SplitRatio, y=ARTime, col sep=comma] {images/f16rat10.csv};
    \addlegendentry{$|S|=599$}

    \addplot[thick, line width=2pt, gray]
      table [x=SplitRatio, y=ARTime, col sep=comma] {images/f16rat5.csv};
    \addlegendentry{$|S|=630$}

    \addplot[thick, line width=2pt, purple]
      table [x=SplitRatio, y=ARTime, col sep=comma] {images/f16rat6.csv};
    \addlegendentry{$|S|=2511$}

    \addplot[thick, line width=2pt, yellow]
      table [x=SplitRatio, y=ARTime, col sep=comma] {images/f16rat7.csv};
    \addlegendentry{$|S|=3405$}

    \addplot[thick, line width=2pt, Green]
      table [x=SplitRatio, y=ARTime, col sep=comma] {images/f16rat3.csv};
    \addlegendentry{$|S|=4175$}

    \addplot[thick, line width=2pt, olive]
      table [x=SplitRatio, y=ARTime, col sep=comma] {images/f16rat9.csv};
    \addlegendentry{$|S|=13842$}

  \end{axis}
\end{tikzpicture}}
    \caption{F-16 First Scenario}
    \label{fig:f16-1-rat}
  \end{subfigure}\hfill
  \begin{subfigure}[t]{0.48\textwidth}
    \centering
    \resizebox{\textwidth}{!}{\begin{tikzpicture}
  \begin{axis}[
      width=4\textwidth,
      height=4\textwidth,
      xlabel={\Huge \bfseries Split Ratio},
      ylabel={\Huge \bfseries  Time (s)},
      xtick={0,0.2,0.4,0.6,0.8,1},
      xlabel style={yshift=-1cm},
      ylabel style={yshift=1cm},
      xticklabel style={font=\Huge}, 
      yticklabel style={font=\Huge}, 
      ymode=log,
      ymajorgrids=true,
      yminorgrids=true,
      minor y tick num=4,              % 4 minor ticks per major step
      major grid style={gray!30},
      minor grid style={gray!10, dashed}, 
      legend style={
        at={(0.5,-0.15)},
        anchor=north,
        legend columns=5,
        font=\Huge,
        draw=black,
        inner sep=2pt
      },
    ]

    \addplot[thick, line width=2pt, blue]
      table [x=SplitRatio, y=ARTime, col sep=comma] {images/f16-1-1.csv};
    \addlegendentry{$|S|=101$}

     \addplot[thick, line width=2pt, magenta]
      table [x=SplitRatio, y=ARTime, col sep=comma] {images/f16-1-10.csv};
    \addlegendentry{$|S|=187$}

    \addplot[thick, line width=2pt, olive]
      table [x=SplitRatio, y=ARTime, col sep=comma] {images/f16-1-9.csv};
    \addlegendentry{$|S|=299$}

     \addplot[thick, line width=2pt, orange]
      table [x=SplitRatio, y=ARTime, col sep=comma] {images/f16-1-4.csv};
    \addlegendentry{$|S|=403$}

    \addplot[thick, line width=2pt, Green]
      table [x=SplitRatio, y=ARTime, col sep=comma] {images/f16-1-3.csv};
    \addlegendentry{$|S|=417$}

    \addplot[thick, line width=2pt, purple]
      table [x=SplitRatio, y=ARTime, col sep=comma] {images/f16-1-6.csv};
    \addlegendentry{$|S|=425$}

    \addplot[thick, line width=2pt, yellow]
      table [x=SplitRatio, y=ARTime, col sep=comma] {images/f16-1-7.csv};
    \addlegendentry{$|S|=621$}

    \addplot[thick, line width=2pt, gray]
      table [x=SplitRatio, y=ARTime, col sep=comma] {images/f16-1-5.csv};
    \addlegendentry{$|S|=879$}

    \addplot[thick, line width=2pt, cyan]
      table [x=SplitRatio, y=ARTime, col sep=comma] {images/f16-1-8.csv};
    \addlegendentry{$|S|=1285$}

    \addplot[thick, line width=2pt, red]
      table [x=SplitRatio, y=ARTime, col sep=comma] {images/f16-1-2.csv};
    \addlegendentry{$|S|=1925$}

  \end{axis}
\end{tikzpicture}}
    \caption{F-16 Second Scenario}
    \label{fig:figrat}
  \end{subfigure}\hfill
  \caption{Execution Time vs. Splitting Ratio of $\hat{s}_\Delta$}
  \label{fig:all-splitrat}
\end{figure}
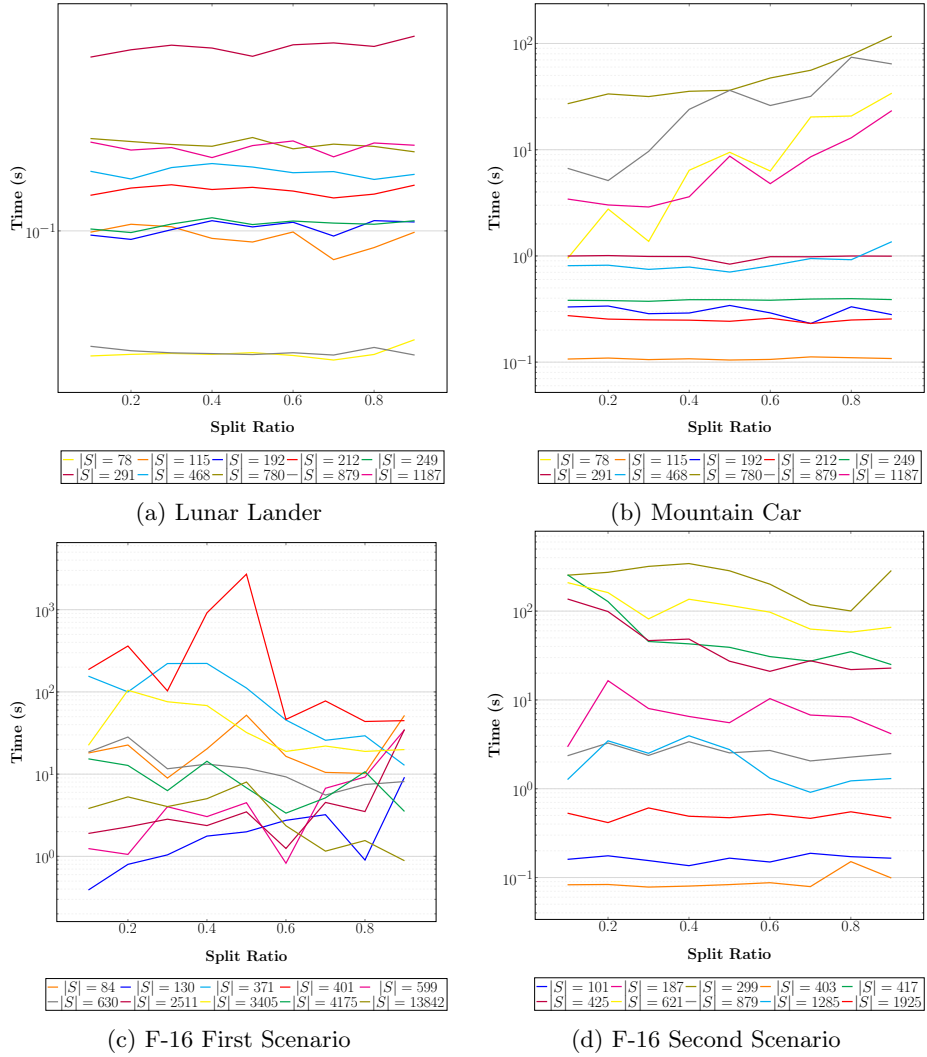

To investigate the impact of the abstraction refinement granularity on overall verification performance, we introduced a tunable hyperparameter \camera{$\beta \in (0, 1]$} which controls the fraction of an abstract state to be preserved when refining the abstract state after the desired property fails to hold in the current abstraction. 
This fraction determines the extent to which the state to be refined is to be partitioned in each refinement iteration, thereby influencing both the model size of the next iteration and the computational overhead.
For each of our four experiment settings, we systematically varied the value of \camera{$\beta$} and recorded the mean execution time across multiple runs on each sample DTMC.
The graphs in \Cref{fig:all-splitrat} illustrate the results of this tuning process.
For example, in \Cref{fig:f16-1-rat} which corresponds to the first scenario of the F-16 Simulator, there is a noticeable improvement in performance across most samples at \camera{$\beta = 0.6$}, making this value a strong candidate.
In our evaluation, in some cases high values of \camera{$\beta$} were observed to produce a proliferation of refinement steps: although each individual split incurs only modest overhead, the cumulative effect of many iterations leads to a rapidly growing abstract state space and, in some cases, worse overall runtime than more aggressive settings.
Conversely, very low values of \camera{$\beta$} leave only a small residual abstracted state, which in many cases creates more states than is necessary to find the cause in most cases, resulting in worse performance.

\end{document}